\documentclass[11pt,a4paper]{emulateapj}
\usepackage{Times}
\usepackage{psfrag}
\usepackage{pspicture}
\usepackage{graphicx}
\usepackage{amsmath}

\def\lir{$L_{\rm IR}$}

\def\teff{\ifmmode T_{\rm eff} \else $T_{\mathrm{eff}}$\fi}

\def\ltsima{$\buildrel<\over\sim$}
\def\lsim{\lower.5ex\hbox{\ltsima}}

\newcommand{\ha}{\ifmmode {\rm H}\alpha \else H$\alpha$\fi}
\newcommand{\hb}{\ifmmode {\rm H}\beta \else H$\beta$\fi}
\newcommand{\lya}{\ifmmode {\rm Ly}\alpha \else Ly$\alpha$\fi}

\newcommand{\heii}{He~{\sc ii}}
\newcommand{\Heiiuv}{He~{\sc ii} $\lambda$1640}

\newcommand{\ebv}{\ifmmode E_{\rm B-V} \else $E_{\rm B-V}$\fi}
\newcommand{\av}{\ifmmode A_{\rm V} \else $A_{\rm V}$\fi}

\def\msun{\ifmmode M_{\odot} \else M$_{\odot}$\fi}
\def\msunyr{\ifmmode M_{\odot} {\rm yr}^{-1} \else M$_{\odot}$ yr$^{-1}$\fi}
\def\zsun{\ifmmode Z_{\odot} \else Z$_{\odot}$\fi}

\def\lsun{\ifmmode L_{\odot} \else L$_{\odot}$\fi}

\def\mup{\ifmmode M_{\rm up} \else M$_{\rm up}$\fi}
\def\mlow{\ifmmode M_{\rm low} \else M$_{\rm low}$\fi}

\lefthead{D. Sobral et al.}  \righthead{PopIII in $z=6.6$ Ly$\alpha$ emitters}

\slugcomment{Accepted for publication in The Astrophysical Journal, June 4, 2015}

\begin{document}

\title{Evidence for PopIII-like stellar populations in the most luminous Lyman-$\alpha$ emitters at the epoch of re-ionisation: spectroscopic confirmation \footnote{Based on observations obtained with X-SHOOTER, FORS2 and SINFONI on the VLT, ESO DDT time (294.A-5018, 294.A-5039) and with DEIMOS on Keck II (U082D).} }

\author{David Sobral\altaffilmark{1,2,3}, Jorryt Matthee\altaffilmark{3}, Behnam Darvish\altaffilmark{4}, Daniel Schaerer\altaffilmark{5,6}, Bahram Mobasher\altaffilmark{4}, \\ Huub J.\ A.\ R\"{o}ttgering\altaffilmark{3}, S\'ergio Santos\altaffilmark{1,2}, Shoubaneh Hemmati\altaffilmark{4}
}
\altaffiltext{1}{Instituto de Astrof\'{\i}sica e Ci\^{e}ncias do Espa\c{c}o, Universidade de Lisboa, OAL, Tapada da Ajuda, PT1349-018 Lisbon, Portugal; FCT-IF/Veni Fellow. E-mail: sobral@iastro.pt}
\altaffiltext{2}{Departamento de F\'{i}sica, Faculdade de Ci\^{e}ncias, Universidade de Lisboa, Edif\'{i}cio C8, Campo Grande, PT1749-016 Lisbon, Portugal}
\altaffiltext{3}{Leiden Observatory, Leiden University, P.O.\ Box 9513, NL-2300 RA Leiden, The Netherlands}
\altaffiltext{4}{Department of Physics and Astronomy, University of California, 900 University Ave., Riverside, CA 92521, USA}
\altaffiltext{5}{Observatoire de Gen\`eve, D\'epartement d'Astronomie, Universit\'e de Gen\`eve, 51 Ch. des Maillettes, 1290 Versoix, Switzerland}
\altaffiltext{6}{CNRS, IRAP, 14 Avenue E. Belin, 31400 Toulouse, France}

\begin{abstract}
\noindent Faint Lyman-$\alpha$ (Ly$\alpha$) emitters become increasingly rarer towards the re-ionisation epoch ($z\sim6-7$). However, observations from a very large ($\sim 5$ deg$^2$) Ly$\alpha$ narrow-band survey at $z=6.6$ \citep{Matthee2015} show that this is not the case for the most luminous emitters, capable of ionising their own local bubbles. Here we present follow-up observations of the two most luminous Ly$\alpha$ candidates in the COSMOS field: `MASOSA' and `CR7'. We used X-SHOOTER, SINFONI and FORS2 on the VLT, and DEIMOS on Keck, to confirm both candidates beyond any doubt. We find redshifts of $z=6.541$ and $z=6.604$ for `MASOSA' and  `CR7', respectively. MASOSA has a strong detection in Ly$\alpha$ with a line width of $386\pm30$\,km\,s$^{-1}$ (FWHM) and with very high EW$_0$ ($>200$\AA), but undetected in the continuum, implying very low stellar mass and a likely young, metal-poor stellar population. `CR7', with an observed Ly$\alpha$ luminosity of $10^{43.92\pm0.05}$\,erg\,s$^{-1}$ is the most luminous Ly$\alpha$ emitter ever found at $z>6$ and is spatially extended ($\sim16$\,kpc). `CR7' reveals a narrow Ly$\alpha$ line with $266\pm15$\,km\,s$^{-1}$ FWHM, being detected in the NIR (rest-frame UV; $\beta=-2.3\pm0.1$) and in IRAC/Spitzer. We detect a narrow He{\sc ii}1640\AA \ emission line ($6\sigma$, FWHM\,$=130\pm30$\,km\,s$^{-1}$) in CR7 which can explain the clear excess seen in the $J$ band photometry (EW$_0\sim80$\,\AA). We find no other emission lines from the UV to the NIR in our X-SHOOTER spectra (He{\sc ii}/O{\sc iii}]1663\AA \ $>3$ and He{\sc ii}/C{\sc iii}]1908\AA \ $>2.5$). We conclude that CR7 is best explained by a combination of a PopIII-like population which dominates the rest-frame UV and the nebular emission and a more normal stellar population which presumably dominates the mass. {\it HST}/WFC3 observations show that the light is indeed spatially separated between a very blue component, coincident with Ly$\alpha$ and He{\sc ii} emission, and two red components ($\sim5$\,kpc away), which dominate the mass. Our findings are consistent with theoretical predictions of a PopIII wave, with PopIII star formation migrating away from the original sites of star formation.
\end{abstract}


\section{Introduction}

The study of the most distant sources such as galaxies, quasars and gamma ray bursts offers unique constraints on early galaxy and structure formation. Such observations are particularly important to test and refine models of galaxy formation and evolution \citep[e.g.][]{Vogelsberger2014,Schaye2015} and to study the epoch of re-ionisation \citep[e.g.][]{Shapiro1994,Sokasian2004,Furlanetto2004,McQuinn2006,Iliev2006}. Over the last two decades, considerable effort has been dedicated towards finding the most distant sources. More recently, and particularly due to the upgraded capabilities of {\it HST}, multiple candidate galaxies up to $z\sim8-11$ \citep[e.g.][]{Bouwens2011,Ellis2013} have been found with deep broad-band photometry. However, spectroscopic confirmation is still limited to a handful of galaxies and quasars at $z > 6.5$ \citep[e.g.][]{Mortlock11,Ono2012,Schenker2014,Finkelstein2013,Pentericci2014,Oesch2015}, for both physical (galaxies becoming increasingly fainter) and observational reasons (the need for deep near-infrared exposures). At these redshifts ($z>6.5$), the Ly$\alpha$ line is virtually the only line available to confirm sources with current instruments. However, Ly$\alpha$ is easily attenuated by dust and neutral hydrogen in the inter-stellar and inter-galactic medium. Indeed, spectroscopic follow-up of UV-selected galaxies indicate that Ly$\alpha$ is suppressed at $z>7$ \citep[e.g.][]{Caruana2014,Tilvi2014} and not a single $z>8$ Ly$\alpha$ emitter candidate has been confirmed yet \citep[e.g.][]{Sobral2009,Faisst2014,Matthee2014}. If the suppression of Ly$\alpha$ is mostly caused by the increase of neutral hydrogen fraction towards higher redshifts, it is clear that $z\sim6.5$ (just over 0.8\,Gyrs after the Big Bang) is a crucial period, because re-ionisation should be close to complete at that redshift \citep[e.g.][]{Fan2006}.

Narrow-band searches have been successful in detecting and confirming Ly$\alpha$ emitters at $z\sim3-7$ \citep[e.g.][]{Cowie98,Malhotra2004,Iye2006,Murayama2007,Hu2010,Ouchi2010}. The results show that the Ly$\alpha$ luminosity function is constant from $z\sim3$ to $z\sim6$, but there are claims that the number density drops from $z\sim6$ to $z\sim6.6$ \citep[e.g.][]{Ouchi2010,Kashikawa2011} and that it drops at an even faster rate up to $z\sim7$ \citep[e.g.][]{Shibuya2012,Konno2014}. Moreover, the fact that the rest-frame UV luminosity function declines from $z\sim3-6$ \cite[e.g.][]{Bouwens2014} while the Ly$\alpha$ luminosity function (LF) is roughly constant over the same redshift range \citep[e.g.][]{Ouchi2008} implies that the cosmic average Ly$\alpha$ escape fraction is likely increasing, from $\sim5$\% at $z\sim2$ \citep[e.g.][]{Hayes2010,Ciardullo2014}, to likely $\sim20-30$\% around $z\sim6$ \citep[e.g.][]{Cassata2015}. Surprisingly, it then seems to fall sharply with increasing redshift beyond $z\sim6.5$. Current results could be a consequence of re-ionisation not being completed at $z\sim6-7$, particularly when taken together with the decline in the fraction of Lyman break selected galaxies with high EW Ly$\alpha$ emission \citep[e.g.][]{Tilvi2014,Caruana2014,Pentericci2014}. However, it is becoming clear that re-ionisation by itself is not enough to explain the rapid decline of the fraction of strong Ly$\alpha$ emitters towards $z\sim7$ \citep[e.g.][]{Dijkstra2014,Mesinger2015}.

It is likely that re-ionisation was very heterogeneous/patchy \citep[e.g.][]{Pentericci2014}, with the early high density regions re-ionising first, followed by the rest of the Universe. If that were the case, this process could have a distinguishable effect on the evolution of the Ly$\alpha$ luminosity function, and it may be that the luminous end of the luminosity function evolves differently from the fainter end, as luminous Ly$\alpha$ emitters should in principle be capable of ionising their surroundings and thus are easier to observe. This is exactly what is found by \cite{Matthee2015}, in agreement with spectroscopic results from \cite{Ono2012}.

In addition to using Ly$\alpha$ emitters to study re-ionisation, they are also useful for identifying the most extreme, metal poor and young galaxies. Studies of Ly$\alpha$ emitters at $z>2-3$ show that, on average, these sources are indeed very metal poor \citep{Finkelstein2011,Nakajima2012,Guaita2013}, presenting high ionisation parameters (high [O{\sc iii}]/H$\beta$ line ratios; \citealt{Nakajima2013}) and very low typical dust extinctions \citep[e.g.][]{Ono2010}. Given these observations, Ly$\alpha$ searches should also be able to find metal-free, PopIII stellar populations (since galaxies dominated by PopIII emit large amounts of Ly$\alpha$ photons, e.g.\,\citealt{Schaerer2002,Schaerer2003}). However, so far, although some candidates for PopIII stellar populations have been found \cite[e.g.][]{Jimenez2006,Dijkstra2007,Nagao2008,Kashikawa2012,Cassata2013}, and some metal poor galaxies have been confirmed \citep[e.g.][]{Prescott2009}, they are all significantly more metal rich than the expected PopIII stars, and show e.g. C{\sc iii}] and C{\sc iv} emission. For example, when there is no evidence for the presence of an AGN and no metal lines, the short-lived He{\sc ii}1640\AA \ emission line (the `smoking gun' for PopIII stars in extremely high EW Lyman-$\alpha$ emitters without any metal emission line) was never detected with high enough EW \citep[e.g.][]{Nagao2008,Kashikawa2012}.

Until recently, Ly$\alpha$ studies at the epoch of re-ionisation have been restricted to the more numerous, relatively faint sources of $L_{\rm Ly\alpha}\sim10^{42.5}$\,erg\,s$^{-1}$ (with some exceptions, e.g. $z=5.7$ follow-up: \citealt{Westra06}; \citealt{Lidman2012}; and `Himiko': \citealt{Ouchi2009}). However, with the wide-field capabilities of current instruments (including Hyper Suprime-Cam; \citealt{HSC2012}), the identification of luminous Ly$\alpha$ emitters will become increasingly easier. Recently, significant progress was made towards finding luminous Ly$\alpha$ emitters at $z=6.6$ \citep{Matthee2015}, through a $\sim5$\,deg$^2$ narrow-band survey, which resulted in the identification of the most luminous Ly$\alpha$ emitters at the epoch of re-ionisation. \cite{Matthee2015} reproduced the Ly$\alpha$ luminosity function of \cite{Ouchi2010} for relatively faint Ly$\alpha$ emitters at $z=6.6$ for the UDS field, who find a decrease in their number density compared to lower redshifts. However, \cite{Matthee2015} find that the luminous end of the $z=6.6$ LF resembles the $z=3-5.7$ luminosity function, and is thus consistent with no evolution at the bright end since $z\sim3$. Extremely luminous Ly$\alpha$ emitters at $z\sim6.6$ are thus found to be much more common than expected, with space densities of $1.5^{+1.2}_{-0.7}\times10^{-5}$ Mpc$^{-3}$. The results may mean that, because such bright sources can be observed at $z\sim6.6$, we are witnessing preferential re-ionisation happening around the most luminous sources first. Such luminous sources may already be free (in their immediate surroundings) of a significant amount of neutral hydrogen, thus making their Ly$\alpha$ emission observable. Furthermore, these sources open a new window towards exploring the stellar populations of the most luminous Ly$\alpha$ emitters at the epoch of re-ionisation even before the {\it James Webb Space Telescope} ({\it JWST}) and $\sim30-40$\,m class telescopes (Extremely Large Telescopes, ELTs) become operational, as these are bright enough to be studied in unprecedented detail with e.g. {\it HST}, ALMA, VLT, Keck.

Here we present spectroscopy of the two most luminous Ly$\alpha$ emitters found so far at the epoch of re-ionisation ($z\sim7$). This paper is organised in the following way. \S2 presents the observations, the Ly$\alpha$ emitter sample, and the data reduction. \S3 outlines the details of the optical and near-infrared spectroscopic observations and measurements with the VLT and Keck data. \S4 presents the discovery of the most luminous Ly$\alpha$ emitters and comparison with previous studies. \S5 discusses SED fitting, model assumptions and how {\it HST} high spatial resolution data corroborates our best interpretation of the data. \S6 presents the discussion of the results. Finally, \S7 outlines the conclusions. A H$_0=70$\,km\,s$^{-1}$\,Mpc$^{-1}$, $\Omega_M=0.3$ and $\Omega_{\Lambda}=0.7$ cosmology is used. We use a Salpeter \citep{Salpeter1955} IMF and all magnitudes are in the AB system, unless noted otherwise.

%
%
%
\begin{table*}
\begin{center}
\caption{Sources and observation log}
\begin{tabular}{ccccccc}
\hline
	Source\tablenote{Observation log for the two luminous Ly$\alpha$ emitter candidates, observed with both the VLT (using X-SHOOTER and SINFONI: CR7 and FORS2: MASOSA) and Keck (using DEIMOS for both sources) and which we have spectroscopically confirmed.} & R.A. & Dec.  & Int. time VLT\tablenote{Note that for X-SHOOTER (CR7) the VIS (visible) and NIR (near-infrared) arms provide different total exposure times (due to different read-out times), and we thus provide them separately in different lines. Keck/DEIMOS exposure times are presented between square brackets: [exposure time].} [Keck] & Dates of observations & Features detected in spectrum\tablenote{We also note the features and lines detected in the spectra: for CR7 we detect rest-frame UV just redder of the Lyman-limit, with a clear Ly$\alpha$ forest, but such flux completely disappears towards redder wavelengths, implying a very blue spectra. No continuum is detected apart from that. An extremely luminous, high EW and narrow Ly$\alpha$ emission line is detected for each of the sources. For CR7, where we have NIR coverage, from both X-SHOOTER and SINFONI, we detect high EW He{\sc ii}1640\AA \ emission fro both instruments, but no other lines.} \\
 & (J2000) & (J2000) & (ks/pixel) &  &   \\ \hline
CR7 & $10\,00\,58.005$ & $+\,01\,48\,15.251$  & VIS: 8.1, [5.4] & 28 Dec 2014; 22 Jan, 15 Feb 2015 & 916-1017\AA, Ly$\alpha$, He{\sc ii}1640\AA  \\
CR7 & $10\,00\,58.005$ & $+\,01\,48\,15.251$  & NIR: 9.9 & 22 Jan, 15 Feb 2015 & He{\sc ii}1640\AA  \\
CR7 & $10\,00\,58.005$ & $+\,01\,48\,15.251$  & SINFONI: 13.5 & 8, 11--13, 17 Mar 2015 and 4 Apr 2015 & He{\sc ii}1640\AA  \\  
MASOSA & $10\,01\,24.801$ & $+\,02\,31\,45.340$& VIS: 6, [2.7] & 29 Dec 2014; 12 Jan 2015 & Ly$\alpha$\\ \hline
\end{tabular}

\end{center}
\label{tab:observations}
\end{table*}

\section{Sample and Spectroscopic Observations}

\subsection{The luminous Ly$\alpha$ candidates at $z=6.6$ }

\cite{Matthee2015} used the Subaru telescope and the NB921 filter on Suprime-cam \citep{Miyazaki2002} to survey $\sim3$\,deg$^2$ in the SA22 (PI: D. Sobral), $\sim1$\,deg$^2$ in COSMOS/UltraVISTA (PI: M. Ouchi) and $\sim1$\,deg$^2$ in UDS/SXDF (PI: M. Ouchi) fields in order to obtain the largest sample of luminous Ly$\alpha$ emitters at the epoch of re-ionisation.

\begin{figure}
\centering
\includegraphics[width=8cm]{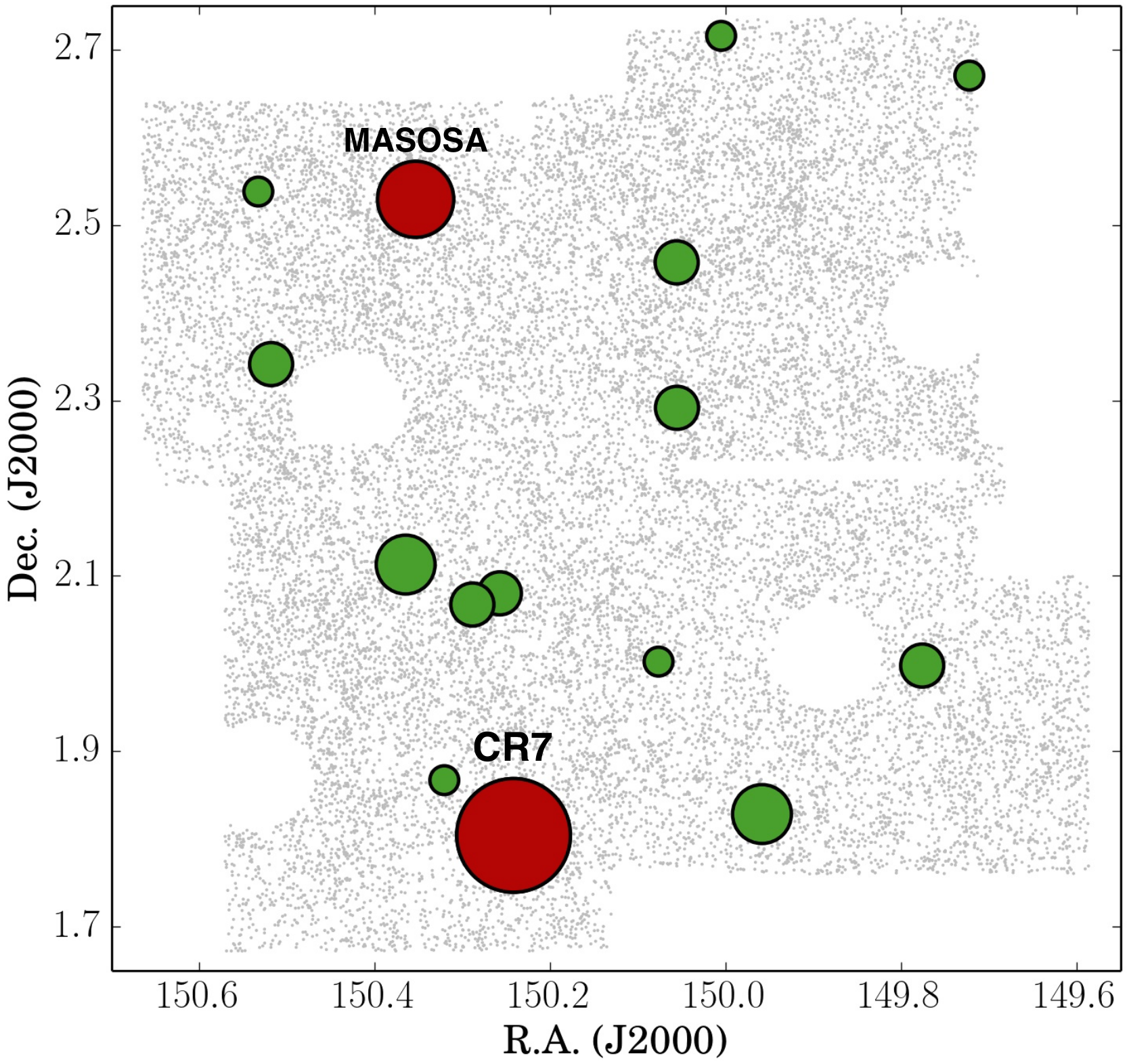}
\caption{Projected positions on sky of all Ly$\alpha$ candidates (green circles) found in the COSMOS/UltraVISTA field. The grey background points represent all detected sources with the NB921 filter, highlighting the masking applied \citep[due to the presence of artefacts caused by bright stars and noisy regions, see][]{Matthee2015}. Ly$\alpha$ candidates are plotted with a symbol size proportional to their Ly$\alpha$ luminosity. CR7 and MASOSA are highlighted in red: these are the most luminous sources found in the field. Their coordinates are given in Table \ref{tab:observations}.}
\label{RADEC}
\end{figure}

Out of the 135 Ly$\alpha$ candidates found in \cite{Matthee2015}, we discover two very bright Ly$\alpha$ candidates in the COSMOS/UltraVISTA field: `CR7' (COSMOS Redshift 7) and MASOSA\footnote{The nickname MASOSA consists of the initials of the first three authors of \cite{Matthee2015}}. MASOSA is particularly compact (0.7\,$''$), while CR7 is extended ($\sim3''$). We show the location of the Ly$\alpha$ emitters within the COSMOS field footprint in Figure \ref{RADEC}, in which the size of the symbols scales with luminosity. We also show their properties in Table \ref{photometry_measurements}.

\begin{figure*}
\centering
\includegraphics[width=16cm]{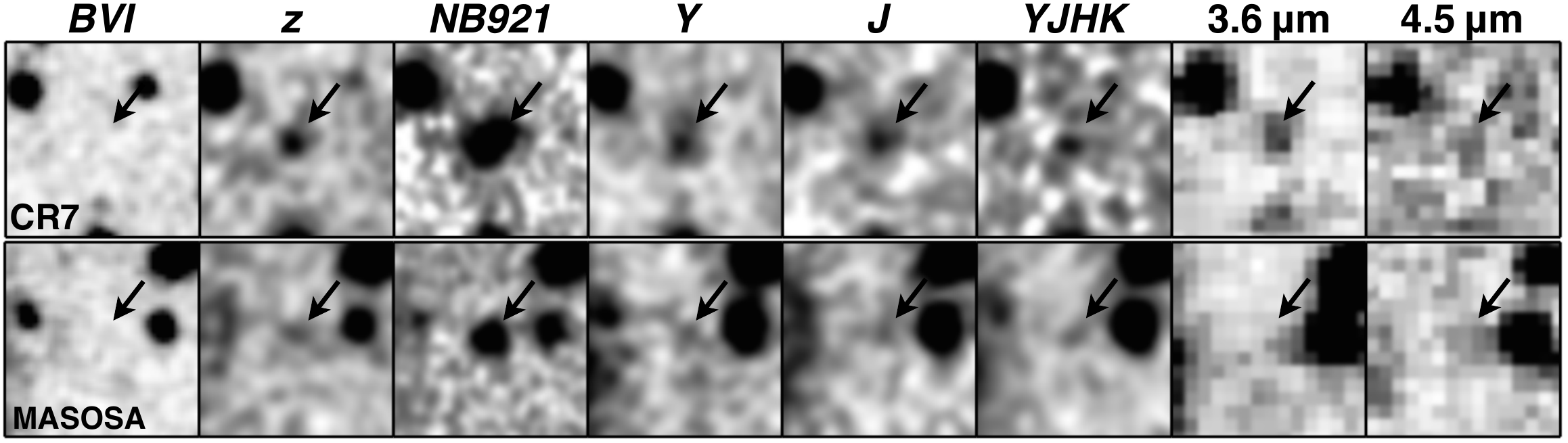}
\caption{Thumbnails of both luminous Ly$\alpha$ emitters in the optical to MIR from left to right. Each thumbnail is $8\times8''$, corresponding to $\sim44\times44$\,kpc at $z\sim6.6$. Note that while for MASOSA the Ly$\alpha$ emission line is detected by the NB921 filter at full transmission, for CR7 the Ly$\alpha$ is only detected at $\sim50$\% transmission. Therefore, the NB921 only captures $\sim50$\% of the Ly$\alpha$ flux: the observed flux coming from the source is $\sim2\times$ larger.}
\label{thumbs}
\end{figure*}

Thumbnails in various wavelengths ranging from observed optical to observed mid-infrared are shown in Figure \ref{thumbs}. Both candidates show very high rest-frame Ly$\alpha$ EWs\footnote{EWs computed by using either $z$ or $Y$ lead to results in excess of 200\,\AA. We also present EWs computed based on $Y$ band and from our spectroscopic follow-up in Table \ref{photometry_measurements}.} in excess of $>$\,200\,\AA. By taking advantage of the wealth of data in the COSMOS/UltraVISTA field \citep[e.g.][]{Capak2007,Scoville2007,Ilbert2009,McCracken2012}, we obtain multi-band photometry for both sources. The measurements are given in Table \ref{photometry_measurements}. We re-measure the 3.6\,$\mu$m and 4.5\,$\mu$m photometry for CR7, in order to remove contamination from a nearby source. Such contamination is at the level of 10-20\%, and is added in quadrature to the photometry errors.

%
%
%
%

We find that CR7 has very clear detections in the near-infrared and mid-infrared (Figure \ref{thumbs}), showing a robust Lyman-break \citep{Steidel1996}. CR7 is detected in IRAC, with colors as expected for a $z\sim6.6$ source \citep{Smit2014}, likely due to contribution from strong nebular lines with EWs in excess of a few 100\,\AA \ in the rest-frame optical (see Figure \ref{thumbs}). Because of the detections in the NIR and MIR, the rest-frame UV counterpart of CR7 was already identified as a $z\sim6-7$ Lyman-break candidate \citep[][]{Bowler2012,Bowler2014}. However, because of its very uncommon NIR colors (i.e. excess in $J$ relative to $Y$, $H$ and $K$), the clear IRAC detections, and, particularly, without the NB921 data (Figure \ref{thumbs}), it was classed as an unreliable candidate, possibly a potential interloper or cool star. MASOSA has a clear detection in the narrow-band and is weakly detected in $z$, but the $z$ band detection can be fully explained by Ly$\alpha$. It is not detected at $>1\sigma$ in the NIR ($J>25.7$, $H>24.5$, $K>24.4$), although a very weak signal is visible in the thumbnails (Figure \ref{thumbs}). This indicates that the Ly$\alpha$ EW is very high and highlights that the Lyman-break selection can easily miss such sources, even if they are extremely bright and compact in Ly$\alpha$, as MASOSA is not detected at the current depth of the UltraVISTA survey \citep{Bowler2014}.

%
%
%
%

\subsection{Spectroscopic observations and data reduction}

Spectroscopic observations were made with the Very Large Telescope (VLT)\footnote{Observations conducted under ESO DDT programs 294.A-5018 and 294.A-5039; PI: D. Sobral.} using X-SHOOTER and SINFONI (for CR7) and FORS2 (for MASOSA). The choice of X-SHOOTER for CR7 was due to the fact that it was detected in the NIR and showed evidence for excess in the $J$ band, likely indicating strong emission lines -- SINFONI was used to confirm the results and avoid potential biases in the slit choice and inclination. Both sources were observed with DEIMOS on the Keck II telescope (see Table \ref{tab:observations}) as well. Spectra for both sources, obtained with both the VLT and Keck, are shown in Figure \ref{figure:spectrum}, including the spectra obtained by combining both data-sets.

\subsubsection{DEIMOS/Keck observations}

DEIMOS/Keck observations targeted both `CR7' and `MASOSA' in two different masks and two different nights. Observations were conducted on 28 and 29 December 2014. The seeing was $\sim0.5''$ on the first night, when we observed `CR7', and $\sim0.7''$ on the second night, when we observed `MASOSA'. Observations were done under clear conditions with midpoint airmass of $<1.1$ for both sources. We used a central wavelength of 7200\,\AA \ and the 600I grating, with a resolution of 0.65\,\AA\,pix$^{-1}$, which allowed us to probe from 4550\,\AA \ to 9850\,\AA. We used the 0.75$''$ slit.

For CR7, we obtained 4 individual exposures of 1.2\,ks and one exposure of 0.6\,ks, resulting in a total of 5.4\,ks. For MASOSA, we obtained a total of 2.7\,ks. A strong, extended and asymmetric line is clearly seen in every individual 1.2\,ks exposure prior to any data reduction. 

We reduced the data using the DEIMOS {\sc spec2d} pipeline \citep{Cooper12,Newman:2012ta}. The observed spectra were flat-fielded, cosmic-ray-removed, sky-subtracted and wavelength-calibrated on a slit-by-slit basis. We used standard Kr, Xe, Ar and Ne arc lamps for wavelength solution and calibration. No dithering pattern was used for sky subtraction. The pipeline also generates 1D spectrum extraction from the reduced 2D per slit, and we use the optimal extraction algorithm \citep{Horne1986}. This extraction creates a one-dimensional spectrum of the target, containing the summed flux at each wavelength in an optimised window. We also extract the spectrum of both sources with varying apertures and at various positions, in order to take advantage of the fact that the sources are clearly spatially resolved. The final spectrum is shown in Figure \ref{figure:spectrum}.

\subsubsection{FORS2/VLT observations}

FORS2/VLT \citep{Appenzeller1998} observations targeted `MASOSA' and were obtained on 12 January and 11 February 2015. The seeing was 0.7$''$ and observations were done under clear conditions. We obtained individual exposures of 1\,ks and applied 3 different offsets along the slit. In total, we obtained 6\,ks. We used the OG590+32 filter together with the GRIS300I+11 Grism (1.62\,\AA\,pix$^{-1}$) with the 1$''$ slit. Ly$\alpha$ is clearly seen in each individual exposure of 1\,ks.

We use the ESO FORS2 pipeline to reduce the data, along with a combination of Python scripts to combine the 2D and extract the 1D. The steps implements follow a similar procedure to that used for DEIMOS.

\

\subsubsection{X-SHOOTER/VLT observations}

Our X-SHOOTER/VLT \citep{Vernet2011} observations targeted `CR7' and were obtained on 22 January 2015 and 15 February 2015. The seeing varied between 0.8$''$ and 0.9$''$ and observations were done under clear conditions. We obtained individual exposures of 0.27\,ks for the optical arm, while for NIR we used individual exposures of 0.11\,ks. We nodded from an A to a B position, including a small jitter box in order to always expose on different pixels. We used 0.9\,$''$ slits for both the optical and near-infrared arms (resolution of $R\sim7500$ and $R\sim5300$, for the optical and near-infrared arms, respectively). In total, for the X-SHOOTER data, we obtained 8.1\,ks in the optical and 9.9\,ks in the NIR. The differences are driven by the slower read-out time in the optical CCD compared to the NIR detector.

\begin{figure*}
\begin{tabular}{cc}
\includegraphics[height=7.6cm]{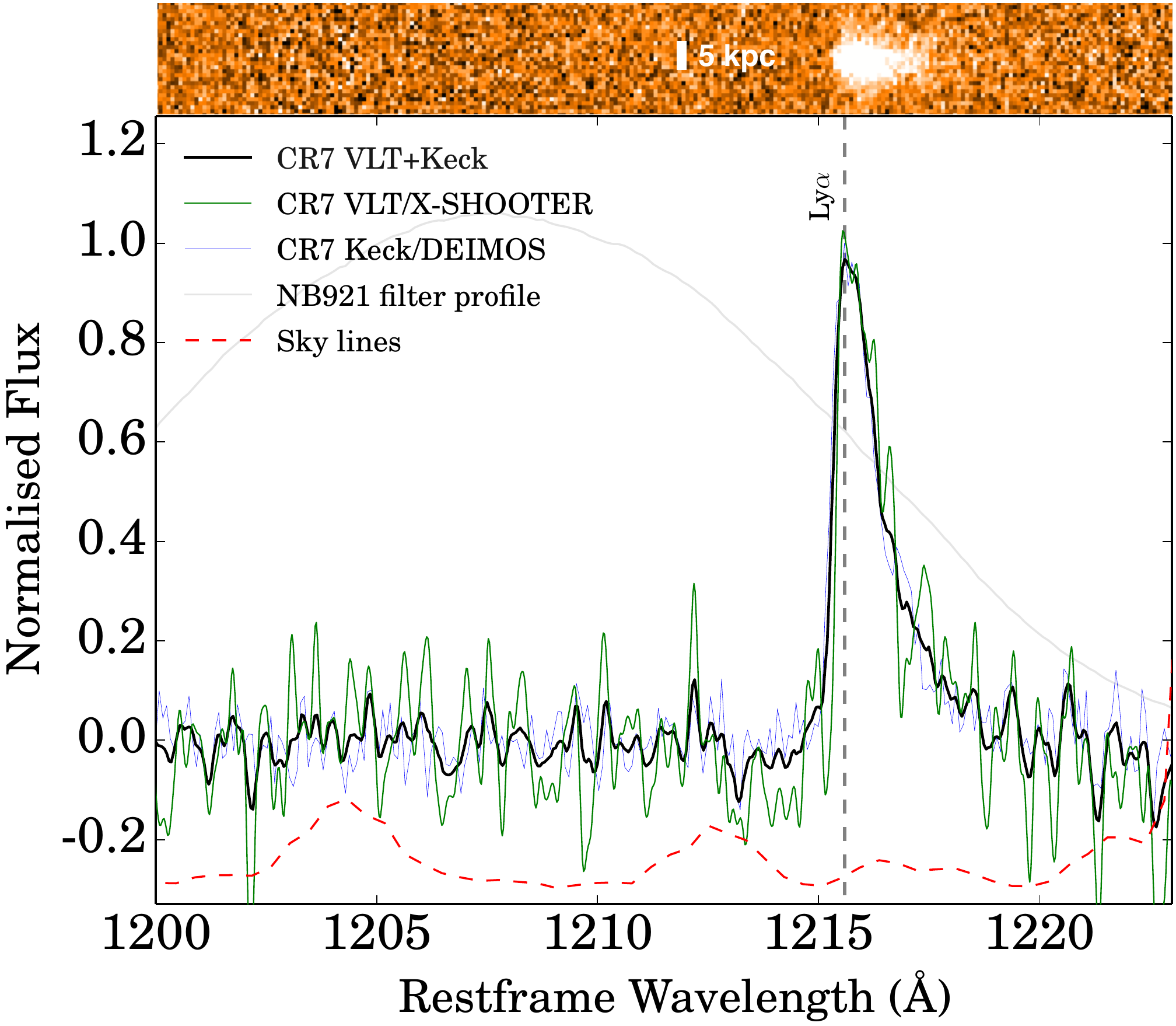}&
\includegraphics[height=7.58cm]{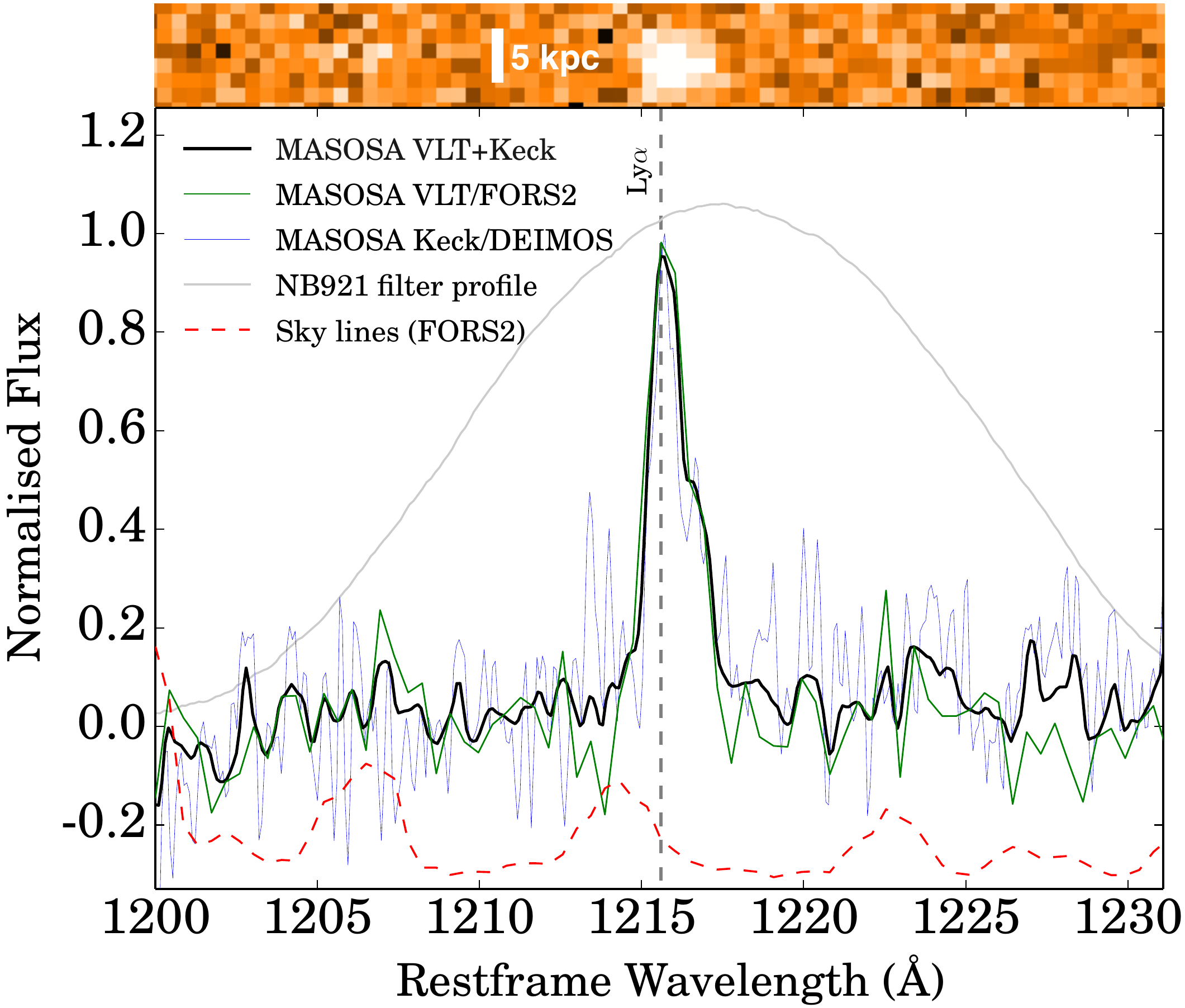}\\
\end{tabular}
\caption{{\it Left}: `CR7' 1-D and 2-D optical spectra, showing the strong and clear Ly$\alpha$ emission line. We also show the NB921 filter profile which was used to select the source. Note that Ly$\alpha$ is detected at the wing of the NB921 filter. Thus, while the NB921 photometry already implied the source was very luminous, its true luminosity was still underestimated by a factor of two. We show both our Keck/DEIMOS and VLT/X-SHOOTER spectra, which show perfect agreement, but with X-SHOOTER providing an even higher spectral resolution, while the DEIMOS spectrum gives an even higher S/N. {\it Right}: `MASOSA' 1-D and 2-D optical spectra (FORS2), showing the strong and clear Ly$\alpha$ emission line. We also show the NB921 filter profile which was used to select the source. We show both the VLT/FORS2 and Keck/DEIMOS spectra, showing they agree very well. The DEIMOS spectrum provides higher resolution, but both clearly reveal the asymmetry of the line, confirming it as Ly$\alpha$ without any doubt.}
\label{figure:spectrum}
\end{figure*}

We use the ESO X-SHOOTER pipeline to fully reduce the visible (optical) and NIR spectra separately. The final spectrum is shown in Figure \ref{figure:spectrum}.

%
%
%
%

\subsubsection{SINFONI/VLT observations}

We have also observed CR7 with the SINFONI \citep{Eisenhauer2003,Bonnet2004} integral field unit on the VLT on 8, 11--13, 17 March 2015 and 4 April 2015. The seeing varied between 0.6$''$ and 0.9$''$ (median: 0.77$''$) in the $J$ band and observations were done under clear conditions. We used the non-adaptive optics mode (spaxel size: 0.25$''$, field of view of $8\times8$$''$) with the $J$ band grism ($R\sim2000$) and individual exposure times of 0.3\,ks. We took advantage of the relatively large spatial coverage to conduct our observations with a jitter box of 2$''$ (9 different positions for each set of 2.7\,ks observations). We obtained 45 exposures of 0.3\,ks each, resulting in a total exposure time of 13.5\,ks.

We use the SINFONI pipeline (v2.5.2) in order to reduce the data. The SINFONI pipeline dark subtracts, extracts the slices, wavelength calibrates, flat-fields and sky-subtracts the data. Flux calibration for each observation was carried out using standard star observations which were taken immediately before or after the science frames. A final stacked data-cube is produced by co-adding reduced data from all the observations. The collapsed data-cube does not result in any continuum detection, as expected given the faint $J$ flux. We extract the 1D spectrum using an aperture of 1$''$.

%
%
%
%
\begin{table}
\caption{A summary of our results for CR7 and MASOSA and comparison to Himiko.}
\begin{center}
\begin{tabular}{cccc}
\hline
Measurement\tablenote{These include both the spectroscopic measurements, but also photometry. In order to provide an easy comparison, we also provide the measurements for Himiko, the other luminous source in the \citet{Matthee2015} sample, fully presented in \citet{Ouchi2013}. Note that the intrinsic Ly$\alpha$ properties may be significantly different from the observed ones, if there is significant absorption by the galaxies themselves.} & CR7 & MASOSA & Himiko  \\
\hline
z$_{spec}$ Ly$\alpha$  & $6.604^{+0.001}_{-0.003}$  & $6.541\pm0.001$ & 6.54  \\    
$\beta$ UV slope & $-2.3\pm0.08$ & ---  & $-2.0\pm0.57$ \\
Ly$\alpha$ (FWHM, km\,s$^{-1}$) & $266\pm15$  & $386\pm30$ & $251\pm21$  \\   
Ly$\alpha$ (EW$_{0,obs,Y}$\tablenote{Using $Y$ band photometry to estimate the continuum.}, \AA) & $211\pm20$ & $>206$ & 78$\pm8$  \\  
Ly$\alpha$ (EW$_{0,obs,spec}$, \AA) & $>230$   & $>200$ & ---  \\  
Ly$\alpha$ (Log$_{10}$L, erg\,s$^{-1}$) & $43.93\pm0.05$ & $43.38\pm0.06$  & $43.40\pm0.07$\tablenote{Re-computed by using $Y$ band to estimate the continuum, in order to match our calculation for CR7 and MASOSA.} \\
Ly$\alpha$/NV & $>70$ & ---  & --- \\   
He{\sc ii}/Ly$\alpha$ & $0.23\pm0.10$ & ---  & --- \\   
He{\sc ii} (EW$_0$, \AA)& $80\pm20$ ($>20$) & ---  & --- \\
He{\sc ii} (FWHM, km\,s$^{-1}$) & $130\pm30$ & ---  & --- \\
He{\sc ii}/OIII]1663 & $>3$ & ---  & --- \\
He{\sc ii}/CIII]1908 & $>2.5$ & ---  & --- \\

\hline
Photometry & CR7 & MASOSA & Himiko \\  
 \hline
$z$ & $25.35\pm0.20$ & $26.28\pm0.37$ & $25.86\pm0.20\tablenote{Measurement from \citet{Ouchi2013}}$\\ 
NB$921$ 2$''$  aperture& $23.70\pm0.04$ & $23.84\pm0.04$ & $23.95\pm0.02$\tablenote{Measurement by \citet{Matthee2015} based on own UDS reduction.} \\
NB$921$ {\sc mag-auto} & $23.24\pm0.03$ & $23.81\pm0.03$ & $23.55\pm0.05^{\rm d}$ \\  
$Y$ & $24.92\pm0.13$ & $>26.35$ & $25.0\pm0.35$\tablenote{Measurement from \citet{Bowler2014} since the $Y$ and $H$ data has been significantly improved with the availability of new VISTA and UKIDSS data.}  \\ 
$J$ & $24.62\pm0.10$ & $>26.15$ & $25.03\pm0.25^{\rm d}$ \\   
$H$ & $25.08\pm0.14$ & $>25.85$ & $25.5\pm0.35^{\rm f}$\\  
$K$ &$25.15\pm0.15$ & $>25.65$ & $24.77\pm0.29^{\rm d}$ \\
3.6$\mu$m & $23.86\pm0.17$ & $>25.6$ & $23.69\pm0.09^{\rm d}$ \\
4.5$\mu$m & $24.52\pm0.61$ & $>25.1$ & $24.28\pm0.19^{\rm d}$ \\

\hline
\end{tabular}
\end{center}
\label{photometry_measurements}
\end{table}

\section{MEASUREMENTS AND SED FITTING}

\subsection{Redshifts}

In both the VLT (FORS2 or X-SHOOTER) and Keck (DEIMOS) spectra for the two targets, we detect the very strong Ly$\alpha$ line (Figure \ref{figure:spectrum}) in emission, and no continuum either directly red-ward or blue-ward of Ly$\alpha$. The very clear asymmetric profiles leave no doubts about them being Ly$\alpha$ and about the secure redshift (Figure \ref{figure:spectrum}). Particularly for CR7, the high S/N $>150\sigma$ (combined Keck and VLT) at Ly$\alpha$, despite the very modest exposure time for such a high-redshift galaxy, clearly reveals this source is unique.

Based on Ly$\alpha$, we obtain redshifts of $z=6.604$ for CR7\footnote{CR7 has a redshift of $z=6.600$ based on He{\sc ii}1640\,\AA \ (see \S\ref{NIR}).} and $z=6.541$ for MASOSA. The redshift determination yields the same answer for both our data-sets: X-SHOOTER and DEIMOS, for CR7 and FORS2 and DEIMOS, for MASOSA (see Figure \ref{figure:spectrum}, which shows the agreement). It is worth noting that for CR7 we find that the Ly$\alpha$ emission line is detected in a lower transmission region of the NB921 filter profile ($50$\% of peak transmission). Therefore, the Ly$\alpha$ luminosity of CR7 is higher than estimated from the NB921 photometry, making the source even more luminous than thought.

\subsection{Spectral line measurements}

By fitting a Gaussian profile to the emission lines, we measure the EW (lower limits, as no continuum is detected) and FWHM. Emission line fluxes are obtained by using NB921 and $Y$ photometry (similarly to e.g. \citealt{Ouchi2009,Ouchi2013}), in combination with the NB921 filter profile and the appropriate redshift. We also check that the integrated emission line (without any assumption on the fitting function) provides results which are fully consistent.

For MASOSA, we find no other line in the optical spectrum, and also find no continuum at any wavelength probed (see Figure \ref{figure:spectrum}). For CR7, we find no continuum either directly blue-ward or red-ward of Ly$\alpha$ in the optical spectrum (both in X-SHOOTER and DEIMOS; Figure \ref{figure:spectrum}). However, we make a continuum detection (spatially very compact) in the rest-frame 916-1017\,\AA \ for CR7 (rest-frame Lyman-Werner photons), with clear absorption features corresponding to the Ly$\alpha$ forest. The reddest wavelength for which we can see continuum directly from the spectra corresponds to Ly$\alpha$ at $z=5.3$. For higher redshifts, the flux is consistent with zero for our spectra. This clear continuum detection at wavelengths slightly redder than the Lyman-limit, but then disappearing for longer wavelengths, can be explained by a combination of a very blue, strong continuum (intense Lyman-Werner radiation) and an average increase of the neutral Hydrogen fraction along the line of sight towards higher redshift, similar to the Gunn-Peterson trough observed in quasar spectra \citep[e.g.][]{Becker2001,Meiksin2005}. However, due to the average transmission of the IGM, the fact that even just a fraction of the light is able to reach us is a unique finding. These findings will be investigated separately in greater detail in a future paper, including further follow-up which will allow an even higher S/N.

\subsection{NIR Spectra of CR7: He{\sc ii} and no other lines} \label{NIR}

We explore our X-SHOOTER NIR spectra to look for any other emission lines in the spectrum of CR7. The photometry reveals a clear $J$ band excess ($0.4\pm0.13$\,mag brighter than expected from $Y$, $H$ and $K$; see Table \ref{photometry_measurements}), which could potentially be explained by strong emission lines (e.g. C{\sc iv}\,1549\,\AA, He{\sc ii}\,1640\,\AA, O{\sc iii}]\,1661\,\AA, O{\sc iii}]\,1666\,\AA, N{\sc iii}]\,1750\,\AA). 

\begin{figure*}
\begin{tabular}{cc}
\includegraphics[height=7.3cm]{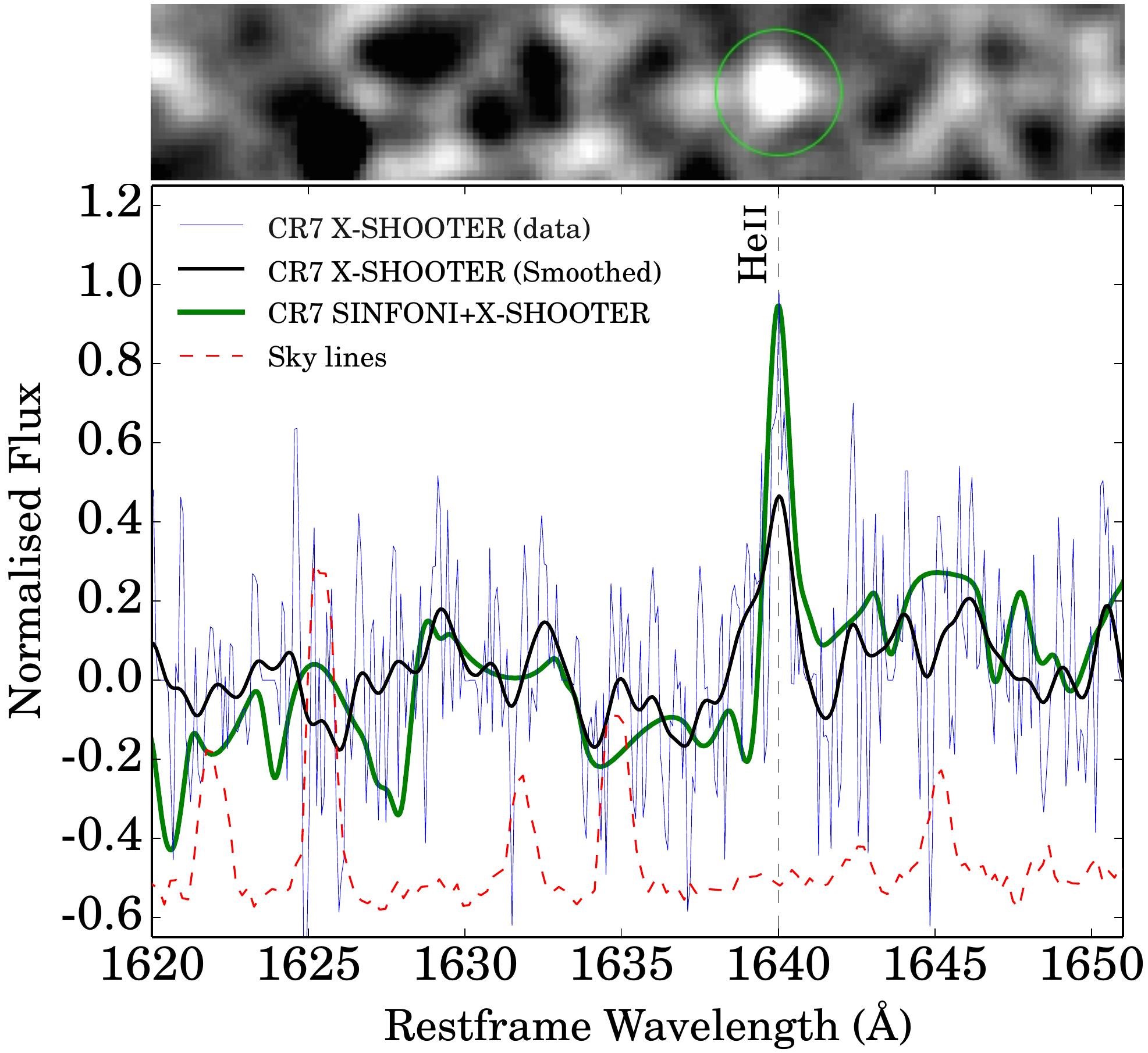}&
\includegraphics[height=7.3cm]{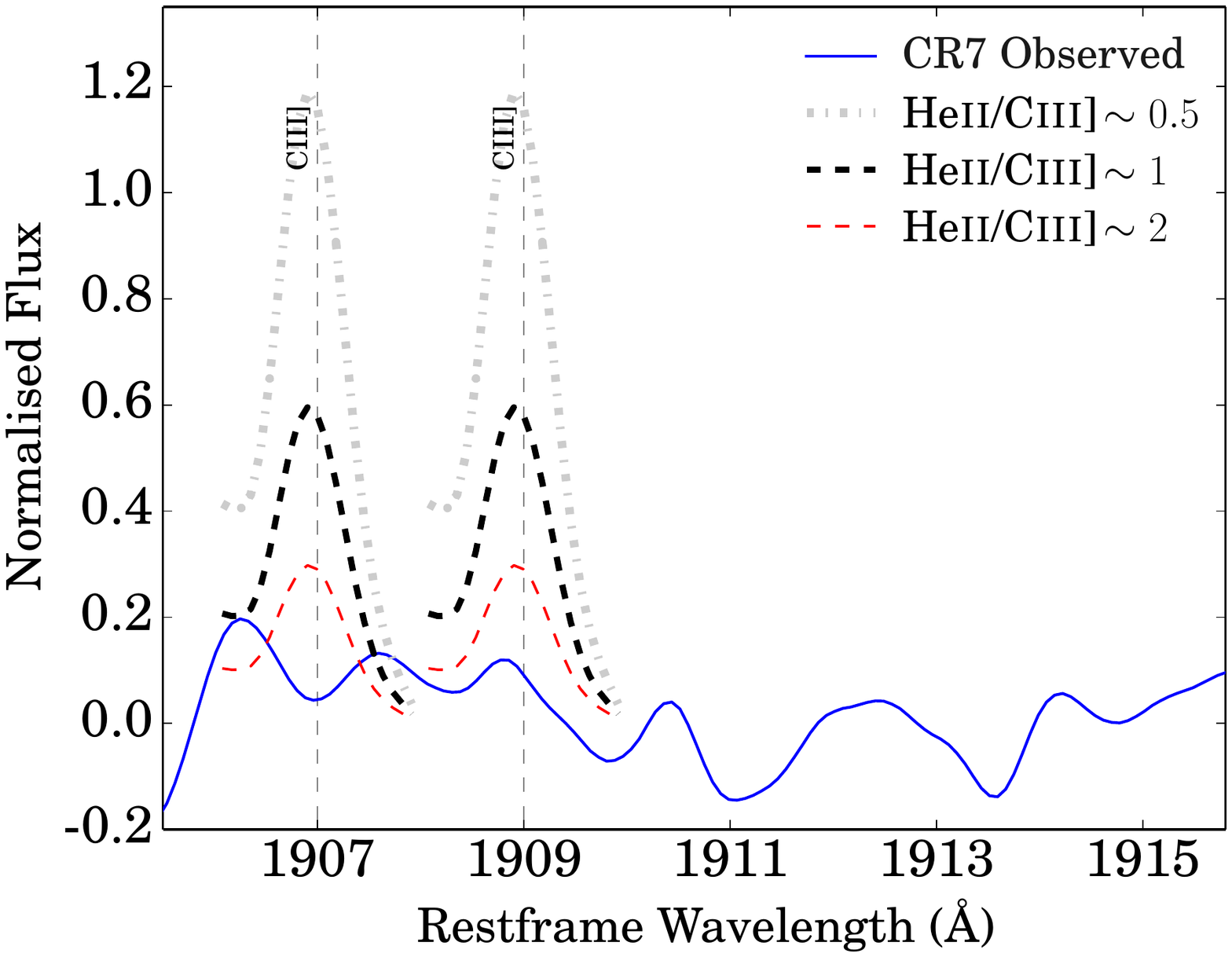}\\
\end{tabular}
\caption{{\it Left:} Our X-SHOOTER NIR spectrum of CR7, revealing a significant detection of the He{\sc ii}1640\,\AA \ emission line. We show both the data at full resolution and binning in wavelength (with a resolution of $0.4$\,\AA). We also show the combined X-SHOOTER and SINFONI data and also show the sky spectrum. We note that sky lines were explicitly masked, but some residuals are still visible, including those of two OH lines just redder of the He{\sc ii} emission line, which show up as a slight flux increase and another one bluer of He{\sc ii}, which has been slightly over-subtracted. We note that no continuum is found in the NIR spectra, and that we would require significantly deeper observations in order to detect it simply based on the NIR photometry. {\it Right:} We investigate the NIR X-SHOOTER spectrum for other emission lines \citep[e.g.][]{Stark2014}. We do not find any other emission line apart from He{\sc ii}, but we show one of the lines that should be stronger in our spectrum (typically $\sim2\times$ stronger than He{\sc ii}, thus He{\sc ii}/C{\sc iii}]$\sim0.5$). We use our HeII line detection and show it at the position of the C{\sc iii}] doublet for a typical line ratio of He{\sc ii}/C{\sc iii}] of 0.5, but also a line ratio of $\sim1$ and $\sim2$. Our data allow us to place a limit of He{\sc ii}/C{\sc iii}]$>2.5$. We also investigate the presence of NV1240, NIV1487, CIV1549, O{\sc iii}]1661, O{\sc iii}]1666, N{\sc iii}]1750: all these are undetected.}
\label{HeII_and_others}
\end{figure*}

We mask all regions for which the error spectrum is too large ($>1.5\times$ the error on OH line free regions), including the strongest OH lines. We then inspect the spectrum for any emission lines. We find an emission line at $12464$\,\AA \ (see Figure \ref{HeII_and_others}). We find no other emission lines in the spectrum (Figure \ref{HeII_and_others}). The emission line found, at $z=6.600$\footnote{The redshift measured from the He{\sc ii}1640\,\AA \ emission line implies a small positive velocity offset between the peak of Ly$\alpha$ and He{\sc ii}1640\,\AA \ of +160\,km\,s$^{-1}$, i.e., the Ly$\alpha$ peak is redshifted by +160\,km\,s$^{-1}$ in respect to He{\sc ii}.}, corresponds to 1640\,\AA, and thus we associate it with He{\sc ii}. Given the line flux ($4.1\pm0.7\times10^{-17}$\,erg\,s$^{-1}$\,cm$^{-2}$), and the level of continuum estimated from e.g. $Y$ and $H$ bands, the line flux we measure is sufficient to explain the $J$ band excess. It also means that we detect He{\sc ii} with a high rest-frame EW ($>20$\AA) with our X-SHOOTER spectra, consistent with the necessary rest-frame EW in order to produce the excess in the $J$ band ($\sim80$\,\AA, reliably estimated from photometry). We find that the He{\sc ii}1640\,\AA \ emission line is narrow ($130\pm30$\,km\,s$^{-1}$ FWHM), and detected at 6\,$\sigma$ in our X-SHOOTER data. Results are presented in Table \ref{photometry_measurements}.

In order to further confirm the reality, strength and flux of the He{\sc ii}1640\,\AA \ line, we also observed CR7 with SINFONI on the VLT. SINFONI reveals an emission line at the central position of CR7 (at the peak of Ly$\alpha$ emission), spatially compact (unresolved) and found at $12464$\,\AA, thus matching our X-SHOOTER results. We co-add the X-SHOOTER and SINFONI spectra and show the results in Figure \ref{HeII_and_others}. When co-adding the spectra, we normalise both spectra at the peak value of He{\sc ii}1640\,\AA \ and explicitly mask the strong OH lines.

Our He{\sc ii}1640\,\AA \ emission line implies a small velocity offset between the peak of Ly$\alpha$ and He{\sc ii}1640\,\AA \ of +160\,km\,s$^{-1}$ (i.e., the Ly$\alpha$ peak is redshifted by +160\,km\,s$^{-1}$ in respect to He{\sc ii}, which could also be interpreted as an outflow). This means that, not surprisingly, we are only detecting the red wing of the Ly$\alpha$ line, while both the blue wing and any potential Ly$\alpha$ component with $<+160$\,km\,s$^{-1}$ is being likely absorbed and cannot be observed. This may imply that the intrinsic Ly$\alpha$ flux or luminosity and the EW will be even larger than measured.

%
%
%

\section{Discovery of the Most Luminous Ly$\alpha$ Emitters}

\subsection{MASOSA} \label{lf}

MASOSA is particularly compact (0.7\,$''$ in diameter, corresponding to 3.8\,kpc diameter at $z=6.541$). It is similar to sources now found by MUSE \citep[e.g.][]{Bacon2014,Karman2015}, that are completely undetected in the continuum and thus similar to typical Ly$\alpha$ emitters that have been found over the last years with e.g. Subaru/Suprime-cam. However, MASOSA is extremely bright in Ly$\alpha$ and thus separates it from the very faint sources found with MUSE in the HDF South and from other Ly$\alpha$ emitters. MASOSA is undetected in all continuum bands at all wavelengths and even the weak detection in $z'$ can be fully explained by the luminous Ly$\alpha$ line and little to no continuum. The estimated rest-frame EW of the Ly$\alpha$ line from spectroscopy is very high ($>$\,200\,\AA), thus implying a likely very metal-poor stellar population \citep{Malhotra2002}. Its nature is likely similar to other metal-poor Ly$\alpha$ emitters \citep{Nagao2008,Ono2010,Ono2012}. Given the very high equivalent width (EW) and no continuum detection, MASOSA is likely extremely young, metal-poor and likely contains low stellar mass ($<10^9$\,M$_{\odot}$).

The Ly$\alpha$ emission line shows a FWHM of $386\pm30$\,km\,s$^{-1}$, and with tentative evidence for two components in the Ly$\alpha$ line \citep[similar to what is found by][]{Ouchi2010}, potentially indicating a merger or radiative transfer effects \citep[e.g.][]{Verhamme2006}. However, with the current ground-base imaging, and the faintness in the continuum, it is not possible to conclude anything about the potential merging nature of the source -- only {\it HST} follow-up can investigate this. However, what is already clear is that MASOSA provides a new example of a relatively compact Ly$\alpha$ emitter with a similar luminosity to Himiko \citep[][]{Ouchi2009,Ouchi2013}, but with a much more extreme EW, revealing that such high luminosity sources present diverse properties and may have a diverse nature.

\subsection{CR7}\label{props}


CR7 clearly stands out as the most luminous Ly$\alpha$ emitter at $z\sim7$, with a luminosity of L$_{Ly\alpha}=10^{43.93\pm0.05}$\,erg\,s$^{-1}$, $\sim3\times$ more luminous than any known Ly$\alpha$ emitter within the epoch of re-ionisation. It also presents a very high rest-frame EW of $>200$\,\AA, and thus a likely intrinsic EW which is even higher because of absorption by the ISM. Our measurements are presented in Table \ref{photometry_measurements}.

CR7 is spatially extended: 3$''$ in diameter, corresponding to $\sim15$\,kpc at $z=6.6$, as seen from its narrow-band image, but also in the spectra. Both the X-SHOOTER and DEIMOS data confirm its spatial extent, with both of them agreeing perfectly on the redshift, extent and FWHM. While the Ly$\alpha$ line profile is narrow (FWHM$\approx270$\,km\,s$^{-1}$), and particularly for such high luminosity, we see evidence for potentially 2 or 3 components (double peaked Ly$\alpha$ emission and a redshifted component towards the South of the source) and/or signs of absorption (see Figure \ref{figure:spectrum}), which indicate a complex dynamical structure. It may also mean that the actual intrinsic FWHM is even narrower. However, such tentative evidence requires confirmation with further spectroscopy obtained over different angles, and particularly by exploring deep imaging with high enough spatial resolution with e.g. {\it HST}.

\subsection{Comparison with Himiko}

We find that CR7 may be seen as similar to Himiko (but much brighter in Ly$\alpha$ and much higher EW) due to both sources presenting a spatial extent of about 3$''$ in diameter. Both could therefore be tentatively classed as Ly$\alpha$ ``blobs'' \citep[e.g.][]{Steidel2000,Matsuda2004,Steidel2011}. However, we note that Himiko is detected at peak transmission in the NB and the NB imaging in which it is detected is 1 mag deeper \citep[see][]{Matthee2015} than the imaging used for the discovery of CR7. While the CR7 Ly$\alpha$ line profile is very narrow, it consists of 2 or potentially 3 components, which may indicate that the source is a double or triple merger, likely similar to Himiko in that respect as well (see \S\ref{HST_imaging} which shows this is very likely the case for CR7). However, a simpler explanation is radiation transfer, which can easily cause such bumps \citep[e.g.][]{Vanzella2010}. There are other similarities to Himiko, including: detections in NIR and a blue IRAC color. CR7 is however a factor $\sim3$ brighter in Ly$\alpha$ emission and has an excess in $J$-band (attributed to He{\sc ii} emission). CR7 is also bluer ($\beta=-2.3\pm0.08$, either using $Y-H$ or $H-K$, following equation 1 of \citealt{Ono2010}) in the rest-frame UV when compared to Himiko (which shows $\beta\sim-2.0$, but note that Himiko shows a red colour from $H$ to $K$ which would imply $\beta\sim0.2$ if those bands are used). Furthermore, while for CR7 we see some rest-frame UV light just redder of the Lyman-limit, corresponding to Lyman-Werner radiation (very compact, coincident with the peak of Ly$\alpha$ emission and the HST detection), this is not seen at all for Himiko \citep[][]{Zabl2015}. Also, while CR7 shows a strong He{\sc ii}1640\,\AA \ emission line, no He{\sc ii} is detected in Himiko, even though \cite{Zabl2015} obtained very deep X-SHOOTER data.

MASOSA is quite different. While it has the highest Ly$\alpha$ peak brightness, it is not extended and not detected in NIR or IRAC. Therefore, MASOSA provides also a new class of sources at the epoch of re-ionisation: as luminous as Himiko, but very compact and with no significant rest-frame UV or rest-frame optical detection at the current UltraVISTA depth.

\section{SED fitting and Model Assumptions}

To interpret the photometry/SED of CR7 we exploit the SED-fitting code of \cite{schaerer&debarros2009} and \cite{schaerer&debarros2010}, which is based on a version of the {\em Hyperz} photometric redshift code of \cite{bolzonellaetal2000}, modified to take nebular emission into account. We have explored a variety of spectral templates including those from the GALAXEV synthesis models of \cite{bruzual&charlot2003}, covering different metallicities (solar, \zsun,   to 1/200 \zsun) and star formation histories (bursts, exponentially declining, exponentially rising). A standard IMF with a Salpeter slope from 0.1 to 100 \msun\ is assumed. We refer to these models as ``standard"/"enriched" SED fits or ``standard"/"enriched" models throughout this paper.

In addition, we also use synthetic spectra from metal-free (PopIII) stellar populations assuming different IMFs (Salpeter, top-heavy), taken from \cite{Schaerer2002,Schaerer2003}. Constant SFR or bursts are explored in this case. We also explore SEDs from ``composite" stellar populations, showing a superposition/mix of PopIII and more normal populations.

Nebular emission from continuum processes and emission lines are added to the spectra predicted as described in \cite{schaerer&debarros2009}. Nebular emission from continuum processes and emission lines are proportional to the Lyman continuum photon production. Whereas many emission lines are included in general, we only include H and He lines for the PopIII case \citep[cf.][]{Schaerer2003}. The intergalactic medium (IGM) is treated with the prescription of \citet{Madau95}. Attenuation by dust is described by the Calzetti law \citep{calzettietal2000}.

%
%
%
\begin{figure*}
\begin{tabular}{cc}
\includegraphics[width=8.5cm]{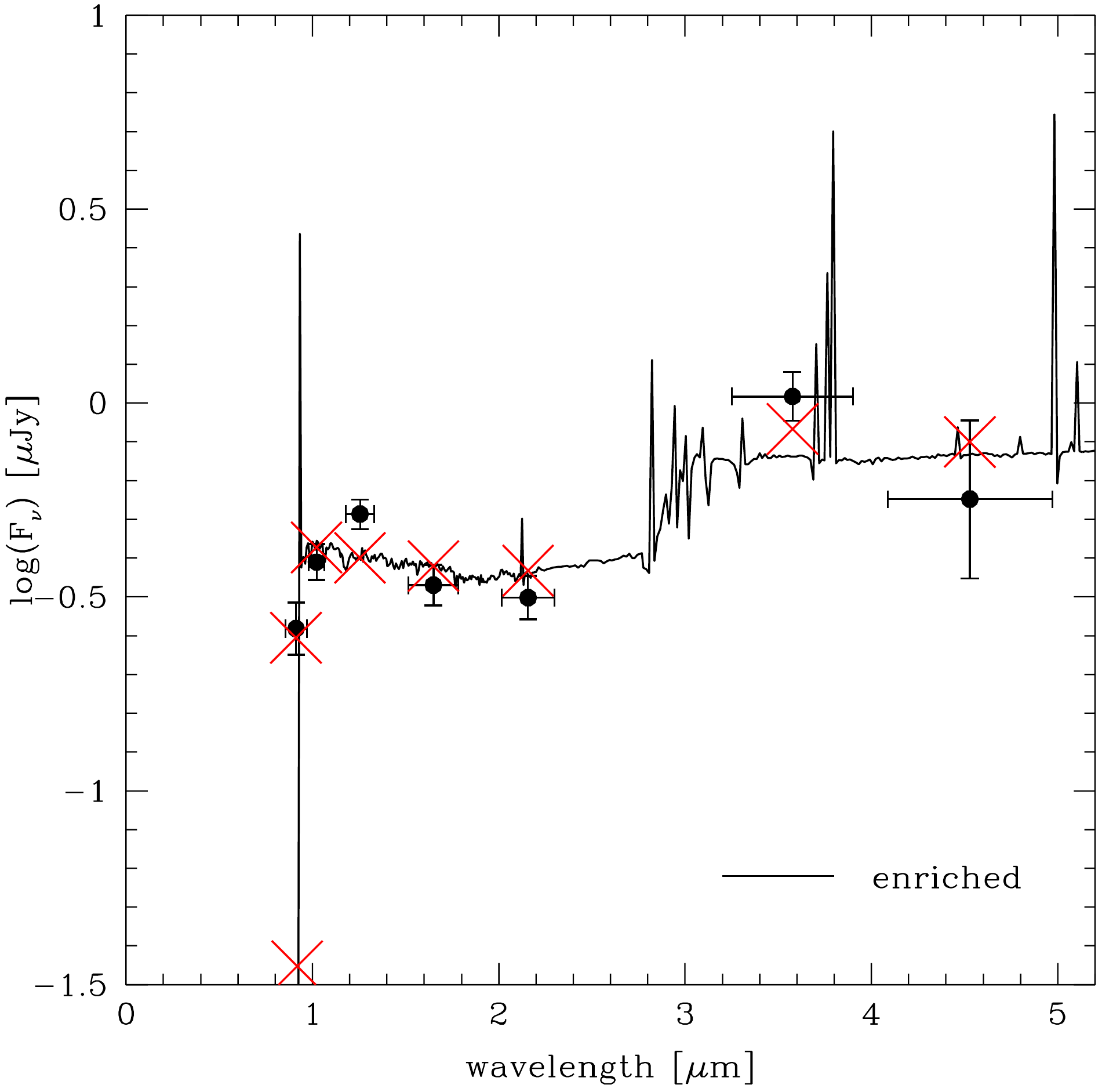}&
\includegraphics[width=8.5cm]{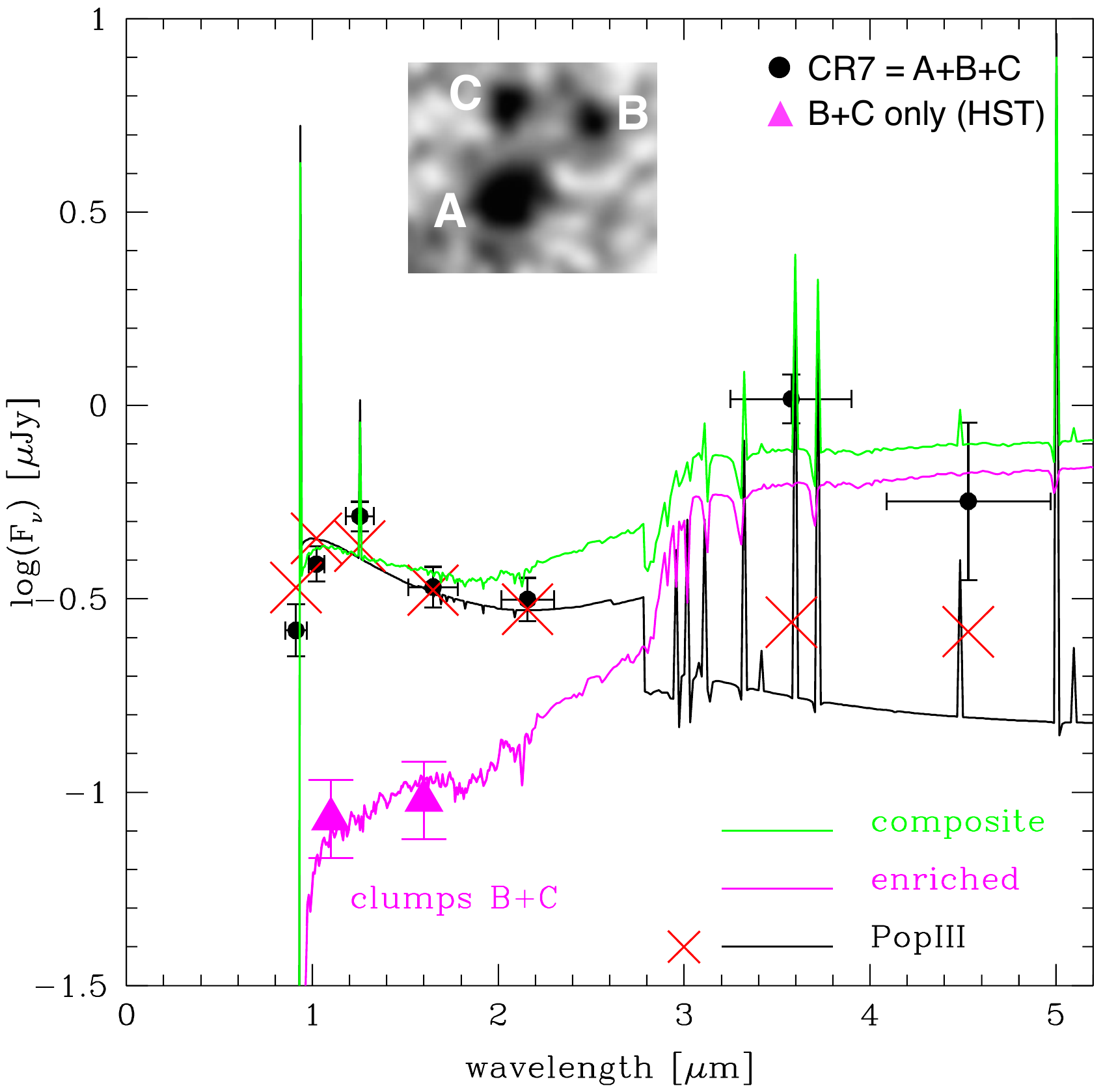}\\
\end{tabular}
\caption{{\it Left}: The SED of `CR7', from observed optical (rest-frame FUV) to observed MIR (rest-frame optical) and the best fit with a with a normal stellar population (not including PopIII stars). Red crosses indicate the flux predicted for each broad-band filter for the best fit. The fit fails to reproduce the strong Ly$\alpha$ emission line and also the excess in $J$ band, due to He{\sc ii} emission. Moreover, and even though the fit is unable to reproduce all the information available for the source, it requires an age of 700\,Myr (the age of the Universe is 800\,Myr at $z=6.6$). In this case, the galaxy would have a SFR of $\sim25$\,M$_{\odot}$\,yr$^{-1}$ and a stellar mass of $\sim10^{10.3}$\,M$_{\odot}$. This, of course, is not able to explain the strong Ly$\alpha$ and the strong He{\sc ii} emission line. {\it Right}: Same observed  SED of ÔCR7Õ as in the left panel plus HST photometry for clumps B+C (magenta triangles). The black line shows a fit with a pure PopIII SED to the rest-frame UV part; the magenta line the SED of an old simple stellar population with 1/5 solar metallicity which matches the flux from clumps B+C; the green line shows the predicted SED summing the two populations after rescaling the PopIII SED by a factor 0.8. The composite SED reproduces well the observed photometry. Although there is a tension between the strength of the He{\sc ii} line and nebular continuum emission (cf. text), a PopIII contribution is required to explain the HeII $\lambda$1640 line and the corresponding excess in the J-band. Although He{\sc ii} is very strong, we find no evidence for any other emission lines that would be characteristic of an AGN. Furthermore, the clear IRAC detections and colors, and particularly when taken as a whole, can be fully explained by a PopIII population, while normal stellar population or AGN is simply not able to.}
\label{SED_FIT}
\end{figure*}

\subsection{CR7: SED fitting with a normal population}\label{props}

Part of the photometry of CR7 is explained relatively well with ``standard" models, as illustrated in Figure \ref{SED_FIT} (left panel). The shortcomings of these fits is that they cannot account for the relative excess in the $J$ band with respect to $Y$, $H$, and $K$, and that the Ly$\alpha$ emission is not strong enough to reproduce the entire flux observed in the NB921 filter. Both of these shortcomings thus relate to the presence of the strong emission lines (Ly$\alpha$ and He{\sc ii}) observed in spectroscopy and also affecting the broad-band photometry.

The typical physical parameters derived from these SED fits which only include ``normal'' stellar populations indicate a stellar mass $M_* \sim 2\times10^{10}$\,M$_{\odot}$, SFR\,$\sim 25$\,M$_{\odot}$\,yr$^{-1}$, and a fairly old age ($\sim 700$ Myr; the Universe is $\sim800$\, Myr old at $z=6.6$). The SED fits and the derived parameters do not vary much for different star-formation histories (SFHs): both for exponentially declining and rising cases, the fits prefer long timescales (i.e. slowly varying SFHs). Whereas for exponentially declining SFHs and for constant SFR the best-fit attenuation is negligible ($A_V=0$), a higher attenuation is needed for rising SFHs, as expected \citep[cf.][]{SP05,Finlator2007}. Depending on the assumed metallicity, this may reach from $A_V=0.5$ (for 1/5 solar) to $0.25$ (for 1/200 \zsun). The corresponding SFR is $\sim 30-40$\,M$_{\odot}$\,yr$^{-1}$, and the predicted IR luminosity ranges from $\log$(\lir / \lsun) = 10.4 to 11.2, assuming energy conservation.
The typical specific SFR, sSFR, obtained from these fits is sSFR $\sim 1.2-1.9$ Gyr$^{-1}$, as expected for a ``mature" stellar population with SFH close to constant \citep[c.f.][]{Gonzalez2014}.

%
%
%
\begin{figure*}
\includegraphics[width=18cm]{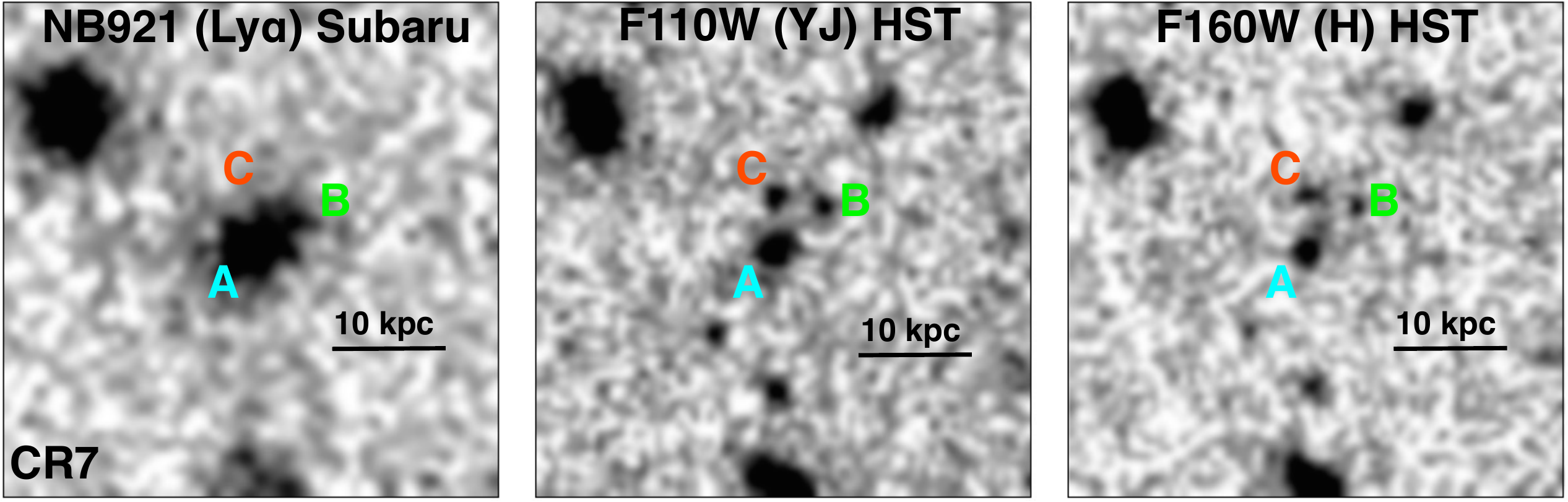}
\caption{{\it Left:} CR7 with the NB921 filter/Suprime-cam imaging on Subaru, showing the extent of the Ly$\alpha$ but note that NB921 detects Ly$\alpha$ at only 50\% transmission. {\it Middle:} HST imaging in $YJ$, revealing that CR7 clearly splits into 3 different components which we name A, B and C. {\it Right:} HST imaging in $H$, again revealing the 3 different components in CR7. We find that component A fully dominates the rest-frame UV and is coincident with the peak of Ly$\alpha$ emission and the location at which we detect strong He{\sc ii}1640\AA \ emission. Clumps B and C are much redder, and fully consistent with significantly contributing to the IRAC photometry. Note that because of the colours of the B and C clumps, they completely dominate the mass of the system, and thus the actual mass centre of the system would be located between C and B, and significantly away from A. This is fully consistent with a scenario in with PopIII star formation is propagated in a wave from the central position towards the outskirts.}
\label{CR7_HST_multiband}
\end{figure*}

\subsection{CR7: SED fitting with contribution from PopIII stars}\label{props}

The presence of strong Ly$\alpha$ and He{\sc ii} emission lines, plus the absence of other UV metal emission lines (cf.\ above), may be due to exceptionally hot stars with a strong and hard ionizing flux, resembling that expected for PopIII stars \cite[cf.][]{Tumlinson2001,Schaerer2002}. A fit with PopIII templates \citep[from][]{Schaerer2002,Schaerer2003} is shown in Figure \ref{SED_FIT} (right panel: black line). PopIII models \citep{Schaerer2002,Schaerer2003} show that at 3.6\,$\mu$m (for $z=6.6$) there are strong He{\sc ii} and He{\sc i} lines, apart from H$\beta$, and that these Helium emission lines should have fluxes comparable to that of H$\beta$ in the case of PopIII \citep{Schaerer2002}. Specifically, 3.6\,$\mu$m should be contaminated by He{\sc ii}4686, He{\sc i}4471 and He{\sc i}5016. For 4.5\,$\mu$m, apart from H$\alpha$, the He{\sc i}5876 should be detected and could be comparable to H$\beta$. The general features of our fits with PopIII templates (considering bursts or constant SFR, as well as different IMFs) are the following: 

\begin{itemize}
\item The UV rest-frame part of the SED is very well reproduced, allowing naturally for a ``boost" of the $J$-band flux due to the presence of the He{\sc ii} line and for a stronger \lya\ line due to a higher Lyman continuum flux (for the same UV flux). The exact strength of these emission lines is very sensitive to the ``details" of the population, such as the upper end of the IMF, the age or star formation history \citep[see more details in e.g.][]{Raiter2010}. The fits with PopIII also reproduce all other NIR detections: {$Y$,} $H$ and $K$ (see Figure \ref{SED_FIT}). 

\item Populations with very strong line-emission also have a strong nebular continuum emission red-ward of \lya\ and increasing towards the Balmer limit \cite[see e.g.][]{Schaerer2002}. The observations ($H$ and $K$ band photometry) do not permit a much stronger contribution from the nebular continuum, as that would mean an increasing redder $H$-$K$ colour, which is not observed. This limits the maximum strength of the predicted emission lines, except if the two emission processes (recombination line emission and nebular continuum emission, which is due to two-photon and free-bound emission) could be decoupled, and emission lines could be increased without significantly increasing the nebular continuum emission.

\item All the available PopIII templates fitting the (rest-)UV part of the spectrum, predict a relatively low flux in the IRAC bands with or without accounting for emission lines. Therefore, a pure metal-free population does not seem to be able to completely reproduce the observed rest frame UV--optical SED of this source. In any case, a PopIII only explanation would not seem very likely. Therefore, a metal-free population alone (without decoupling between recombination line emission and nebular continuum emission) is not able to reproduce the observed rest-frame UV-optical SED of this source.

\end{itemize}

\subsection{PopIII and a more chemically evolved stellar population}\label{props_composite}

As a consequence of our findings, we are led to consider a hybrid SED consisting of two populations (the source may well be a merger, and these components may well be completely separated, avoiding pollution by metals into the potentially metal-free region): a young metal-free stellar population and a more chemically evolved population. This can be fully confirmed by using {\it HST} and appropriate filters that can easily isolate Ly$\alpha$, the rest-frame UV, and He{\sc ii}.

A superposition of a young PopIII component dominating in the UV and an older population of ``normal" metallicity (in this case, 0.2\,\zsun) dominating the rest-frame optical flux of CR7 is shown in the right panel of Figure \ref{SED_FIT}. In practice, we add 80\% of a metal-free simple stellar population with an age of 16 Myr (shown by the black line; although younger ages of $\la 5$ Myr are preferred to produce the strongest He{\sc ii} emission) to a 360 Myr old burst of 1/5 \zsun\ (magenta curve), giving the total flux shown in green.
As can be seen, an ``old'' population of $\sim 1.6 \times 10^{10}$ \msun\ can make up for the missing rest-frame optical flux, whereas a young PopIII burst can dominate the UV and the emission lines. There is some tension/uncertainty in the age or age spread of the metal-free population, as very young ages
($\la 5$ Myr) are preferred to produce the strongest HeII emission, whereas slightly older ages are preferred to avoid too strong nebular continuum emission (cf.\ above). For indication, the mass of the metal-free component would be $1.4 \times 10^{9}$ \msun\ for a Salpeter IMF from 1 to 500 \msun, i.e.\ $\sim 9$ \% of mass of the old population. However, significantly less mass could be needed if the PopIII IMF was flat or top-heavy, lacking e.g.\ completely low mass stars (stars below 10\,M$_{}\odot$). This could mean that $\sim10^7$\,M$_{\odot}$ of PopIII stars would be needed in order to fully explain the flux for an IMF peaking at $\sim60$\,M$_{\odot}$, or even less if the IMF peaks at even higher masses. This reveals that the presence of a young, metal-free population, forming for example in a yet un-polluted region of the galaxy, in an slightly evolved galaxy at $z=6.6$, could reproduce the observed features of CR7 \citep[consistent with theoretical predictions from e.g.][]{Scannapieco2003,Tornatore2007}.

\subsection{{\it HST} imaging of CR7}\label{HST_imaging}

In order for our best interpretation to be valid, CR7 would require to be clearly separated/resolved, with $HST$ resolution, into at least two different spatial components: one being dominated by a PopIII-like stellar population (dominating the UV light but with only a very small fraction of the mass), and another, redder in the UV (and fully dominating the mass), with fluxes similar to those shown in Figure \ref{SED_FIT}. While this requires more detailed follow-up, we find that CR7 has been fortuitously observed and is in the FoV of previous WFC3 observations in F110W (broad $YJ$ filter that also contains Ly$\alpha$) and F160W ($H$) of a different project (ID: 12578, PI: Forster-Schreiber). We explore such data in order to investigate the rest-frame UV morphology of CR7 and to conduct a first study of the rest-frame UV colours. Observations in F110W and F160W were obtained for 2.6\,ks each.

\begin{figure}
\centering
\includegraphics[width=7cm]{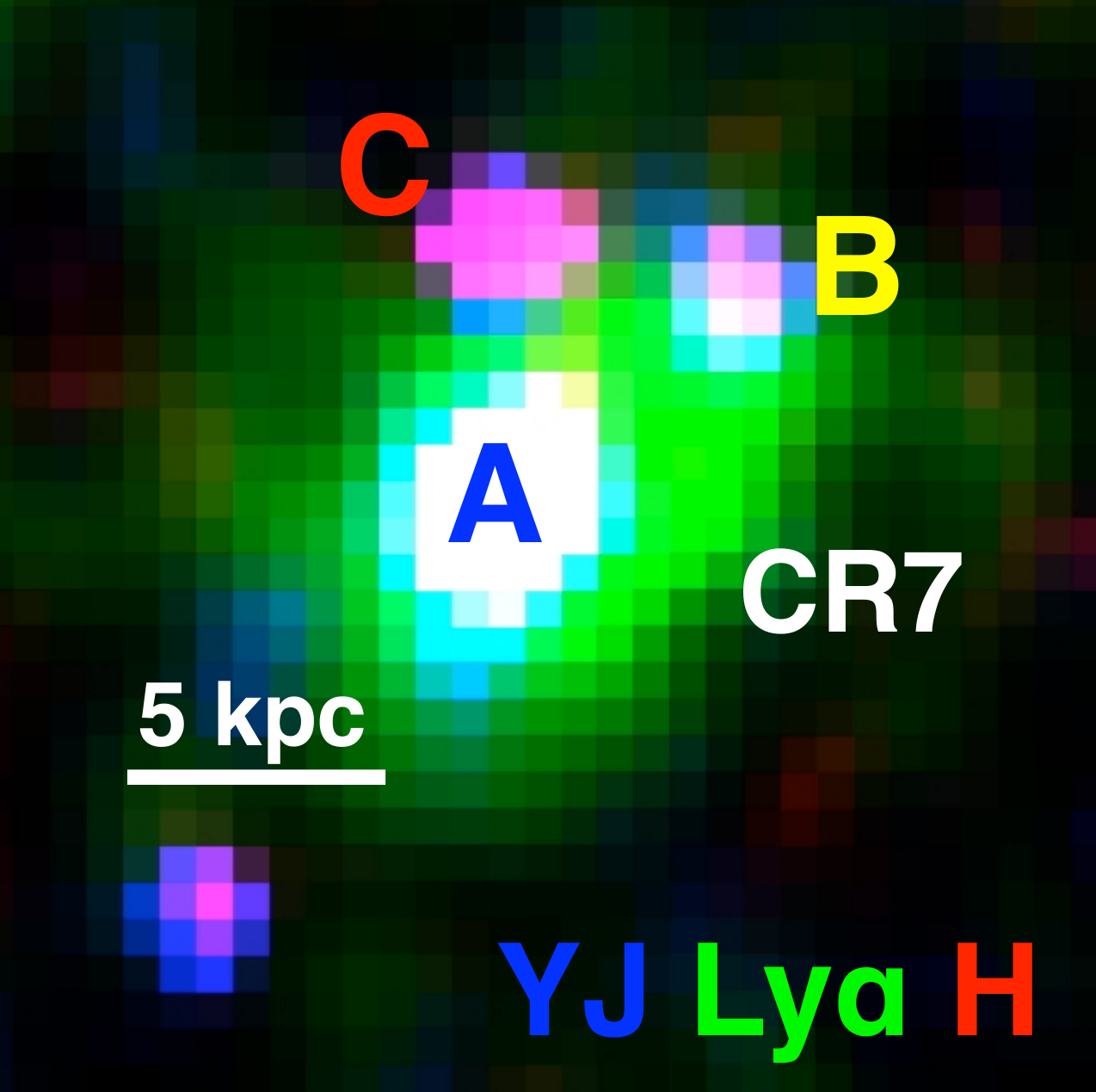}
\caption{A false colour composite of CR7 by using NB921/Suprime-cam imaging (Ly$\alpha$) and two {\it HST}/WFC3 filters: F110W (YJ) and F160W (H). This shows that while component A is the one that dominates the Ly$\alpha$ emission and the rest-frame UV light, the (likely) scattered Ly$\alpha$ emission seems to extend all the way to B and part of C, likely indicating a significant amount of gas in the system. Note that the reddest (in rest-frame UV) clump is C, with B having a more intermediate colour and with A being very blue in the rest-frame UV.}
\label{CR7_HST}
\end{figure}

We show the data in Figure \ref{CR7_HST_multiband}, including a comparison with out NB921 imaging. Figure \ref{CR7_HST} presents a false-colour image combining data obtained with NB921, F110W and F160W. We find that CR7 is, beyond any doubt, split in different components (see Figures \ref{CR7_HST_multiband} and \ref{CR7_HST}), in line with our best interpretation of a PopIII-like stellar population which dominates the UV, and a redder stellar population, found to be physically separated by at least 5\,kpc (projected). In fact, we actually find three different components, which we label as A, B and C (see Figures \ref{CR7_HST_multiband} and \ref{CR7_HST}). We obtain photometry for each of the clumps separately, in order to quantitatively test if they could explain the UV photometry predicted for the two components in \S\ref{props_composite}. We use 0.4$''$ apertures for components B and C and 1$''$ for component A, more spatially extended.

\begin{figure}
\centering
\includegraphics[width=8.25cm]{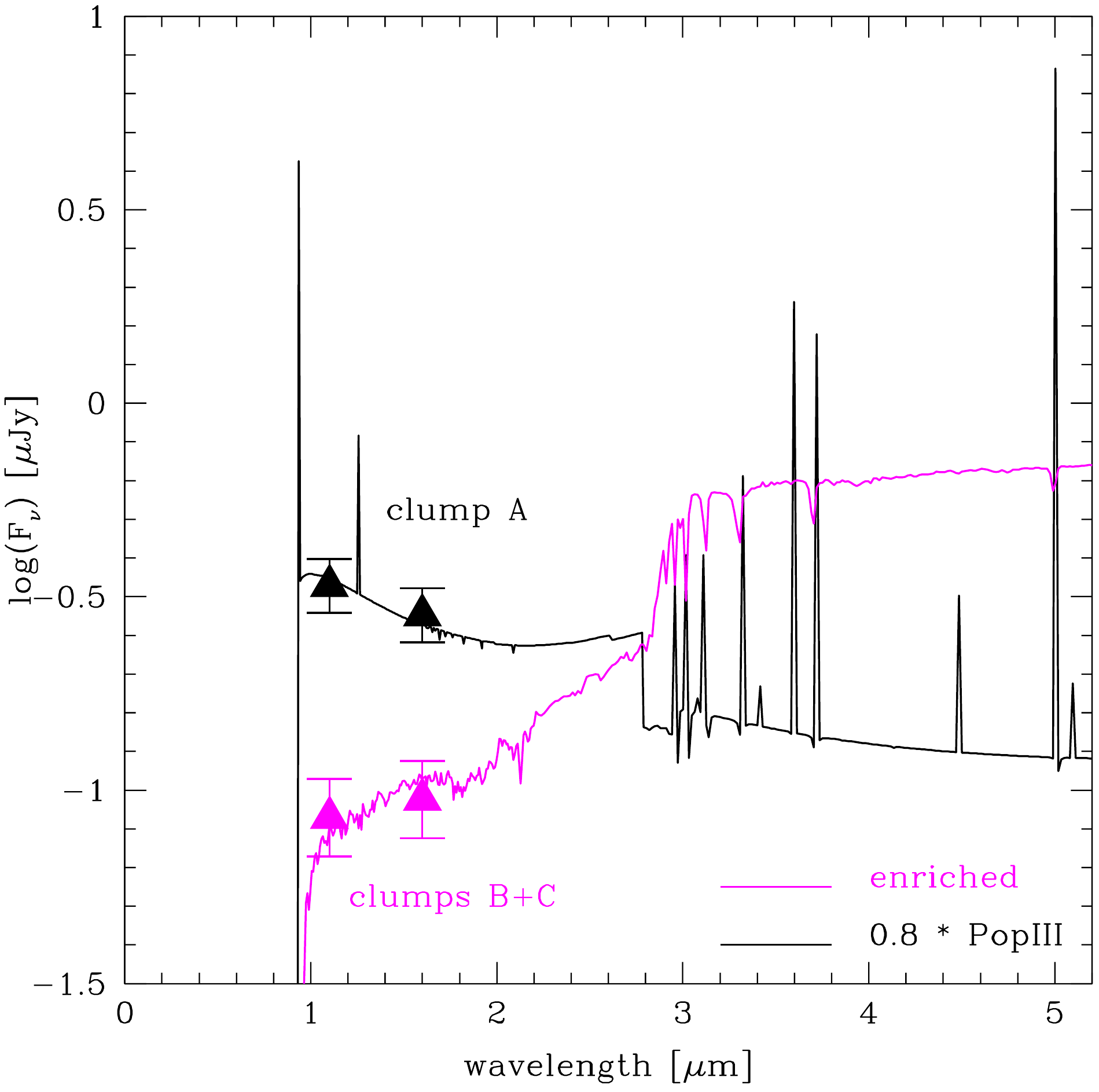}
\caption{{\it HST} imaging in $YJ$ and $H$ allows us the physically separate CR7 in two very different stellar populations and show remarkable agreement with our best-fit composite SED derived in \S\ref{props_composite}. While clump A (see e.g. Figure \ref{CR7_HST}) is very blue and dominates the rest-frame UV flux, B+C are red and likely dominate the rest-frame optical and the mass. Note that the we simply show the {\it HST} data together with our best fit composite model derived in \S\ref{props_composite} which was solely based on the full photometry and did not make use of any resolved {\it HST} data.}
\label{CR7_HST_SED}
\end{figure}

We find that the sum of the two redder, fainter clumps (B+C) matches our evolved stellar population remarkably well (see Figures \ref{SED_FIT} and \ref{CR7_HST_SED}).  Note that the photometry of the two redder clumps was not used to derive such fit. We find that the central clump (A) is the one that dominates the rest-frame UV light (Figure \ref{CR7_HST_SED}). Figure \ref{CR7_HST} also shows the rest-frame colours of components A, B and C. We also find that the peak of He{\sc ii} emission line (from SINFONI, although with low spatial S/N) and the peak of Ly$\alpha$ (high S/N but extended) likely originate (within the errors) from the centre of the brightest UV clump (A). The clumps are physically separated by $\sim5$\,kpc.

The {\it HST} imaging reveals that CR7 may be either a triple merger (similar to Himiko), and/or a system where we are witnessing a PopIII star formation wave, which may have moved from the reddest clump (C) to the other (B) and we are observing the brightest UV clump at the right time (A). We have direct evidence of the intense Lyman-Werner radiation (rest-frame 912-1000\,\AA) from the brightest UV clump. It is therefore possible that the other clumps have emitted as much or even more of such radiation a few $\sim100$\,Myrs before, preventing what is now the site of young massive stars (A) to form before and potentially allowing for that pocket of metal free gas to remain metal free. There are of course, other potential interpretations of our observations. In \S\ref{discussion} we discuss the different potential scenarios in detail.

%
%
%
%

%
%
%
%

\section{DISCUSSION} \label{discussion}

\subsection{The nature of CR7} \label{lf}

CR7, with a luminosity of L$_{Ly\alpha}=10^{43.93\pm0.05}$\,erg\,s$^{-1}$ is $\sim3\times$ more luminous than any known Ly$\alpha$ emitter within the epoch of re-ionisation \citep[e.g.][]{Ouchi2013}.

Our optical spectrum shows that the source is very blue towards the extreme ultra-violet up to the Lyman limit at 912 \AA \ rest-frame, as we detect some faint continuum (spatially very compact) at rest-frame $\sim916-1017$\,\AA \ (Lyman-Werner radiation). X-SHOOTER data also provides a near-infrared spectrum, allowing to investigate the significant excess seen in the $J$ band photometry from UltraVISTA \citep[][]{McCracken2012,Bowler2014}, indicative of emission line(s). No continuum is detected in the NIR spectrum. However, and despite the relatively low integration time, a strong He{\sc ii}1640\,{\AA} line was found ($\sim6\sigma$), capable of explaining the excess in the $J$ band (see Figure \ref{SED_FIT}). He{\sc ii} can only be produced if the intrinsic extreme UV spectrum is very hard, i.e., emits a large number of ionising photons with energies above 54.4\,eV, capable of ionising He completely. From our X-SHOOTER spectra, we place a lower limit in the rest-frame EW of the He{\sc ii} line of $>20$\,\AA, but estimate from photometry that the line has EW$_0=80\pm20$\,\AA, consistent with our spectroscopic lower limit. The line we detect is also narrower than Ly$\alpha$, with FWHM of $130\pm30$\,km\,s$^{-1}$, as He{\sc ii}1640\,\AA \ does not scatter easily as Ly$\alpha$, as the line is not self-resonant.

While in principle there are a variety of processes that could produce both high EW Ly$\alpha$ and He{\sc ii}1640\AA, some of them are very unlikely to produce them at the luminosities we are observing, such as X-ray binaries or shocks. However, in principle, cooling radiation could produce strong Ly$\alpha$ emission with luminosities similar to those measured for CR7. \cite{Faucher2010} provides predictions of the total Ly$\alpha$ cooling luminosity as a function of halo mass and redshift. Under the most optimistic/extreme assumptions, it would be possible to produce a Ly$\alpha$ luminosity of $\sim10^{44}$\,erg\,s$^{-1}$ for a dark matter halo mass of $M>5\times10^{11}$\,M$_{\odot}$. Since such dark matter haloes should have a co-moving number density of about $\sim10^{-5}$\,Mpc$^{-3}$ at $z\sim6.6$, their number densities could potentially match the luminous Ly$\alpha$ emitters that we have found. However, in the case of cooling radiation, the He{\sc ii} emission line should be significantly weaker than what we measure (with intrinsic He{\sc ii}/Ly$\alpha$ of 0.1 at most; e.g. \citealt{Yang2006}), and cooling radiation in a massive dark matter halo should also result in a broader Ly$\alpha$ line than what we observe.

There are, nonetheless, four main sources known to emit an ionising spectrum that can produce high luminosity, high EW nebular Ly$\alpha$ and He{\sc ii} as seen in our spectra \citep[see also similar discussion in e.g.][]{Prescott2009,Cai2011,Kashikawa2012}:

\smallskip

1) strong AGN (many examples have been found, particularly on e.g. radio galaxies: \citealt{DeBreuck2000}), with typical FWHM of lines being $\sim1000$\,km\,s$^{-1}$;

\smallskip

2) Wolf-Rayet (WR) stars (many cases known in e.g. SDSS: \citealt{Shirazi2012}; or see \cite{Erb2010} for a higher redshift example), with typical FWHM of lines being $\sim3000$\,km\,s$^{-1}$;

\smallskip

3) Direct collapse black hole (DCBH), which have been predicted and studied theoretically \citep[e.g.][]{Agarwal2013, Agarwal2015}, although none has been identified yet, but they should produce strong He{\sc ii}1640\AA \ \citep{Johnson2011};

\smallskip

4) PopIII stars \citep[e.g.][]{Schaerer2003,Raiter2010}, which should produce high EW, narrow He{\sc ii} emission lines (FWHM of a few $\sim100$\,km\,s$^{-1}$). 

\smallskip

Many potential candidates for PopIII have been identified based on their colours and/or high EW Ly$\alpha$, but either no He{\sc ii} was found \citep[e.g.][]{Nagao2008}, He{\sc ii} was found but with clear signatures of AGN activity \citep{DeBreuck2000,Matsuoka2009}, or He{\sc ii} had very low EW \citep{Cassata2013}. Thus, so far, not a single source has been found with high EW Ly$\alpha$, He{\sc ii} detection with high EW, no AGN signatures, no WR star signatures (e.g. P-Cygni profiles, broad lines, and many other metal lines e.g. \citealt{Brinchmann2008}; see also \citealt{Grafener2015}) and with no other metal lines.

\subsection{The nature of CR7: AGN or WR stars?} \label{lf}

In order to test the possibility of CR7 being an AGN, we start by checking X-ray data. We find no detection in the X-rays, with a limit of $<10^{44}$\,erg\,s$^{-1}$ \citep{Elvis2009}. We also find no radio emission, although the limit is much less stringent than the X-ray emission. The X-SHOOTER and DEIMOS spectra were carefully investigated for any metal lines, particularly NV, Oxygen and Carbon lines (see Table \ref{photometry_measurements} and also Figure \ref{HeII_and_others}). No such lines were found, and thus we place 1$\sigma$ upper limits on their fluxes, to constrain the nature of the source, finding, e.g. Ly$\alpha$/NV\,$>70$. Our Ly$\alpha$ and He{\sc ii} emission lines are narrow (both FWHM$\sim100-300$\,km\,s$^{1}$), thus excluding broad-line AGN. Narrow-line AGNs with He{\sc ii} emission typically have CIII]1909/He{\sc ii}$\sim1.5\pm0.5$ \citep[e.g.][]{DeBreuck2000}; such line ratio would result in a strong CIII]1909 detection in our X-SHOOTER spectrum (see Figure \ref{HeII_and_others}). We do not detect C{\sc iii}]1909 and obtain a strong upper limit of C{\sc iii}]1909/He{\sc ii}$<0.4$ (1$\sigma$; see Figure \ref{HeII_and_others}), which greatly disfavours the AGN hypothesis. The strong limit on Ly$\alpha$/N{\sc V}\,$>70$ also disfavours the AGN hypothesis and points towards very low metallicities.

There are no indications of WR stars, due to the very narrow He{\sc ii} line ($\sim100$\,km\,s$^{-1}$, compared to typical FWHM of $\sim3000$\,km\,s$^{-1}$ for WR stars, c.f. \citealt{Brinchmann2008}) and no other metal lines.

We note, nonetheless, that while CR7 is strongly disfavoured as an AGN, it shows characteristics of what has been predicted for a direct collapse black hole \citep[e.g.][]{Johnson2011, Agarwal2013, Agarwal2015}. This is because it shows $\beta=-2.3$ (as predicted), no metal lines, and high luminosity. The detection of the other nearby sources about $\sim5$\,kpc away (see Figure \ref{CR7_HST_multiband}) are also a key prediction from Agarwal et al. 2015, while the relatively high observed He{\sc ii}/Ly$\alpha$ would also match predictions for a direct collapse black hole \citep{Johnson2011}. However, CR7 does not show any broad line, as predicted by \cite{Agarwal2013}, and the observed Ly$\alpha$ and He{\sc ii} luminosities are higher by about $\sim2$ orders of magnitude when compared to predictions by e.g. \cite{Johnson2011} for the case of $\sim1-5\times10^4$\,M$_{\odot}$ black holes. Another key distinction between a direct collapse black hole and stellar population(s) is X-ray emission: if it is a black hole, it must be emitting much more X-ray flux than a PopIII stellar population, and thus would likely be detectable with Chandra given the high line luminosities measured. Deeper Chandra observations could in principle test this.

\subsection{On the hardness of the ionizing source of CR7}

The theory of recombination lines relates, to first order, the ratio of \heii\ to Hydrogen recombination lines to the ratio between the ionizing photon flux above 54 eV,  $Q({\rm He^+})$, and that above 13.6 eV, $Q(H)$, the energies needed to ionize He$^+$ and H respectively. For the relative intensity \Heiiuv/\lya\ one thus has:
 \begin{equation} I(1640)/I(Ly\alpha) \approx 0.55 \times \frac{Q({\rm He^+})}{Q(H)},
\end{equation}
where the numerical factor  depends somewhat on the electron temperature, here taken to be $T_e=30$ kK \citep{Schaerer2002}. The observed line ratio \Heiiuv/\lya\ $\approx 0.22$ therefore translates to $Q({\rm He^+})/Q(H) \approx0.42$, which indicates a very hard ionizing spectrum. For metal-free stellar atmospheres such a hardness is only achieved in stars with very high effective temperatures, typically $\teff > 140$ kK (or $>100$ kK for 1 $\sigma$ lower limit), hotter than the (already hot) zero-age main sequence predicted for PopIII stars which asymptotes to $\teff \approx 100$ kK for the most massive stars \citep[cf.][]{Schaerer2002}. For integrated stellar populations consisting of a ensemble of stars of different masses, a maximum hardness $Q({\rm He^+})/Q(H) \approx0.1$ is expected for zero or very low metallicities \citep{Schaerer2002,Schaerer2003}, higher than inferred from the observed  \Heiiuv/\lya\ ratio of CR7. This could indicate that only a fraction of the intrinsic \lya\ emission is observed, or that a source with a spectrum other than predicted by the above PopIII models (e.g.\ an AGN) is responsible for the ionization or contributing at high energies ($>54$ eV). In fact the observed \Heiiuv\ equivalent width of $80 \pm 20$ \AA\ is in good agreement with the maximum equivalent width predicted for PopIII models \citep{Schaerer2002}. This supports the explanation that $\sim$ 75\% of the intrinsic \lya\ emission may have escaped our observation, e.g.\ due to scattering by the IGM, internal absorption by dust, or due to a low surface brightness halo, processes which are known to affect in general \lya\ emission \citep[e.g.][]{Atek2008,Dijkstra2011, Steidel2011}. If the intrinsic \lya\ emission is $\ga 3-4$ times higher than observed, the hardness ratio is compatible with ``standard" PopIII models.
Future observations and detailed photoionization models may yield further insight on the properties of the ionizing source of CR7.

\subsection{CR7: A PopIII-like stellar population?}

As the AGN and WR stars hypothesis are strongly disfavoured (although we note that a direct collapse black hole could still explain most of our observations), could CR7 be dominated by a PopIII-like stellar population? For this to be the case, such stellar population would have to explain the detection, FWHMs, EWs and fluxes of Ly$\alpha$ and He{\sc ii} (including the strong excess in $J$ band), the UV continuum and continuum slope ($\beta=-2.3\pm0.08$) and the IRAC detections which imply very high EW rest-frame optical lines. We have shown in \S\ref{props_composite} and \S\ref{HST_imaging} that a composite of PopIII models \citep{Schaerer2002,Schaerer2003,Raiter2010} with a more evolved stellar population can match all our observations, including spatially resolved {\it HST} data. We note that the intrinsic He{\sc ii}/Ly$\alpha$ line ratio predicted for PopIII would be $\sim0.05-0.1$ \citep[][]{Schaerer2002,Schaerer2003}, but that can easily result in an observable ratio of $\sim0.2-0.3$ if a significant fraction of the Ly$\alpha$ line is absorbed/attenuated by neutral Hydrogen. Since we find a velocity offset between He{\sc ii} and Ly$\alpha$, our observations would support a an intrinsic He{\sc ii}/Ly$\alpha$ line ratio much closer to $\sim0.05-0.1$, thus strongly favouring a PopIII-like population.

A key question, of course, is whether it is even possible or expected to observe Lyman-$\alpha$ coming from PopIII stars alone, even if such line is ultra-luminous. The most massive PopIII stars should be short-lived (a few Myr), and, without any previous contribution to ionise their surroundings from e.g. neighbour star clusters or other nearby proto-galaxies, the most massive PopIII stars would have to be able to emit enough ionising photons to produce an ionised sphere larger than 1\,Mpc after less than a few Myrs \citep[][]{CenHaiman}, before the most massive stars reach the supernovae phase and likely start enriching the local environment. However, such process (for a single PopIII population in full isolation, and fully surrounded by neutral Hydrogen) should take at least $\sim3$\,Myrs to happen: this is simply set by the speed of light. However, if neighbouring sources (either PopIII or PopII stars) have already contributed towards ionising a local bubble, and if PopIII star formation can proceed in a wave-like pattern, likely from the highest density regions to the lowest densities, by the time later PopIII stars form (still in pristine gas which was not contaminated due to being sufficiently far away), they will be in ideal conditions to be directly observed in Ly$\alpha$. Thus, it may be much more likely to observe potentially composite populations than to observe pure PopIII stellar populations. Furthermore, and despite the nature of the stellar populations, it is likely that the observability of very luminous Ly$\alpha$ emitters is strongly favoured in complex systems which already have older stellar populations like CR7 (that were able to ionise local bubbles before and thus allowing for strong Ly$\alpha$ emission from young stellar populations to be observable).

The detection of the signatures from the first generation of stars would provide the first glimpse into the general properties of these stars, such as their initial mass function (IMF). The mass of stars is a fundamental property, as it determines the evolution, the chemical enrichment of their surroundings and their faith. Theoretical work predicts that the first generation of stars are very massive, up to 1000 M$_{\odot}$ \citep{Bromm2004}. More recent work is aiming to begin to sample the potential IMF of PopIII stars, finding a distribution which is very flat or top-heavy, with masses ranging from 10 to 1000\,M$_{\odot}$ \citep[e.g.][]{Hirano2014}, although the PopIII IMF is extremely uncertain \citep[e.g.][]{Greif2011} and only observations will be able to constrain it.

\section{Conclusions}

We presented the spectroscopic follow-up of the two most luminous $z\sim6.6$ Ly$\alpha$ candidates in the COSMOS field ($L_{\rm Ly\alpha}\sim3-9\times10^{43}$\,erg\,s$^{-1}$): `MASOSA' and `CR7'. These sources were identified in \cite{Matthee2015}, revealing that such luminous sources are much more common than previously thought and have number densities of $\sim1.5\times10^{-5}$\,Mpc$^{-3}$. Our main results are:

\begin{itemize}

\item We used X-SHOOTER, SINFONI and FORS2 on the VLT, and DEIMOS on Keck, to confirm both candidates beyond any doubt. We find redshifts of $z=6.604$ and $z=6.541$ for `CR7' and `MASOSA', respectively. `CR7' has an observed Ly$\alpha$ luminosity of $10^{43.93\pm0.05}$\,erg\,s$^{-1}$ ($\sim3\times$ more luminous than Himiko) and is the most luminous Ly$\alpha$ emitter ever found at the epoch of re-ionisation.

\item MASOSA has a strong detection in Ly$\alpha$, with very high Ly$\alpha$ EW (EW$_0>200$\,\AA), implying very low stellar mass and a likely extreme stellar population. It is, nonetheless, undetected in all other available bands and Ly$\alpha$ is also rather compact.

\item CR7, with a narrow Ly$\alpha$ line with 266$\pm15$\,km\,s$^{-1}$ FWHM, is detected in the NIR (rest-frame UV), with $\beta=-2.3\pm0.08$, an excess in $J$ band, and it is strongly detected in IRAC/Spitzer.

\item We detect a strong He{\sc ii}1640\AA \ narrow emission line with both X-SHOOTER and SINFONI in CR7 (implying $z=6.600$), which is enough to explain the clear excess seen in the $J$ band photometry. We find no other emission lines from the UV to the NIR in our X-SHOOTER spectra. No AGN line is seen, nor any signatures of WR stars, as the He{\sc ii}1640\AA \ emission line is narrow (FWHM\,=\,$130\pm30$\,km\,s$^{-1}$). The He{\sc ii}1640\AA \ emission line implies that we are seeing the peak of Ly$\alpha$ emission redshifted by $+160$\,km\,s$^{-1}$ and thus that we are only seeing the red wing of the Ly$\alpha$ (the intrinsic Ly$\alpha$ flux is thus likely much higher than seen), or we are witnessing an outflow.

\item The AGN and WR stars interpretation of the nature of CR7 are strongly disfavoured. An alternative interpretation is that the source hosts a direct collapse black hole, although the lack of broad emission lines and the lack of X-ray detection also disfavours this interpretation. Given all the current data, we conclude that CR7 may host an unseen, extreme stellar population and it is therefore the strongest candidate for PopIII-like stellar population found so far.

\item We find that CR7 cannot be described only by a PopIII stellar population, particularly due to the very strong IRAC detections. Our best interpretation of the full data (spectroscopy and photometry), which is fully consistent with many theoretical predictions, is a combination of a PopIII-like stellar population, which dominates the rest-frame UV and the emission lines, and an older, likely metal enriched stellar population, which is red, and that completely dominates the mass of the system. This interpretation fits remarkably well with high resolution {\it HST}/WFC3 imaging that reveals two red components, each about 5\,kpc away from the peak of the rest-frame UV, Ly$\alpha$ and He{\sc ii}1640\,\AA \ emission and that have the fluxes predicted by our SED fitting.

\end{itemize}

We may be witnessing, for the first time, direct evidence for the occurrence of waves of PopIII-like star formation which could happen from an original star cluster outwards (resulting from strong feedback which can delay PopIII star formation), as suggested by e.g. \cite{Tornatore2007}. In this scenario, the reddest clump in CR7 (C, see Figures \ref{CR7_HST} and \ref{CR7_HST_SED}), which formed first (reddest and oldest), was likely responsible for not only starting to ionise a local bubble, but also for producing copious amounts of Lyman-Werner radiation and feedback that may have prevented star-formation to occur in the vicinity of clump C. Star-formation likely proceeded to the second clump once stellar feedback from C declined (B, see Figure \ref{CR7_HST}), with similar effects (preventing star-formation outside such region, but further ionising a local bubble), and we are observing source A (see Figure \ref{CR7_HST}) at the right time to see an intense PopIII-like star formation episode. Most importantly, this scenario also provides a very simple explanation of why Ly$\alpha$ photons can easily escape the CR7 galaxy, as we see a significant amount of older stars which were able to emit a significant amount of UV photons for a few hundred million years before observations, enough to ionise a bubble of $>1$\,Mpc around the source. Given the strong Lyman-Werner flux that we can infer is coming from the bright UV core, it is very likely that any previous episodes of star-formation could have prevented the gas around those star-forming regions to form stars. Furthermore, we note that while radiation is able to affect the surroundings as it travels fast, significant metal enrichment is very inefficient on scales larger than $\sim1$\,kpc \citep[e.g.][]{Scannapieco2003,Tornatore2007,Ritter2014}, both due to the larger time-scales for metals (from supernovae) to travel outwards (compared to the speed of light), but also due to the continued infall of metal-free gas from the cosmic web. It is therefore likely that in some cases, scales beyond 1-2\,kpc of previous star formation activity can easily have the pristine gas necessary to allow PopIII to form \citep[e.g.][]{Ritter2012,Ritter2014} even at $z\sim6.6$, and to be detectable at redshifts even below $z\sim5$ \citep[e.g.][]{Scannapieco2003,Tornatore2007}. \cite{Tornatore2007}, for example, predict that the peak of PopIII star formation rate density to occur at $z\sim6-8$.

The spectroscopic confirmation of MASOSA and CR7, along with the high S/N spectroscopic and photometric data, allowed us to have a first glimpse into sources likely similar to Himiko and brighter, that are much more common than previously expected and have a remarkable nature. The follow-up of the full \cite{Matthee2015} Ly$\alpha$ sources at even higher luminosities found in the SA22 field will allow us to explore even more the diversity and nature of such unique targets. Such luminous Ly$\alpha$ emitters are the ideal first targets for JWST, particularly due to the likely very high EW and bright optical rest-frame emission lines, which may not be restricted to bright [O{\sc iii}]5007, H$\beta$ and H$\alpha$, but actually include He{\sc ii}4686, He{\sc i}4471, He{\sc i}5016 and He{\sc i}5876. In the case of CR7, a composite of a PopIII-like stellar population and a likely enriched stellar population which is physically separated by $\sim5$\,kpc (Figure \ref{CR7_HST}) is currently strongly favoured, and it is possible that similar stellar populations will be found in the other Ly$\alpha$ emitters. We may have found an actual population with clear signatures of PopIII-like stars, besides CR7. JWST will, in only a short exposure time, clearly show if the rest-frame spectra is made up of only He+H lines, confirming PopIII beyond any doubt, or if [O{\sc iii}] is also present and exactly where each of the lines is coming from.

\section*{Acknowledgments}

We thank the anonymous reviewer for useful and constructive comments and suggestions which greatly improved the quality and clarity of our work. DS acknowledges financial support from the Netherlands Organisation for Scientific research (NWO) through a Veni fellowship, from FCT through a FCT Investigator Starting Grant and Start-up Grant (IF/01154/2012/CP0189/CT0010), from FCT grant UID/FIS/04434/2013, and from LSF and LKBF. JM acknowledges the award of a Huygens PhD fellowship. HR acknowledges support from the ERC Advanced Investigator program NewClusters 321271. The authors would like to thank Mark Dijkstra, Bhaskar Agarwal, Jarrett Johnson, Andrea Ferrara, Jarle Brinchmann, Rebecca Bowler, George Becker, Emma Curtis-Lake, Milos Milosavljevic, Raffaella Schneider, Paul Shapiro and Erik Zackrisson for interesting, stimulating and helpful discussions. The authors are extremely grateful to ESO for the award of ESO DDT time (294.A-5018 \& 294.A-5039) which allowed the spectroscopic confirmation of both sources and the detailed investigation of their nature. Observations are also based on data from W.M. Keck Observatory. The W.M. Keck Observatory is operated as a scientific partnership of Caltech, the University of California and the National Aeronautics and Space Administration. Based on observations obtained with MegaPrime/Megacam, a joint project of CFHT and CEA/IRFU, at the Canada-France-Hawaii Telescope (CFHT) which is operated by the National Research Council (NRC) of Canada, the Institut National des Science de lÕUnivers of the Centre National de la Recherche Scientifique (CNRS) of France, and the University of Hawaii. This work is based in part on data products produced at Terapix available at the Canadian Astronomy Data Centre as part of the Canada-France-Hawaii Telescope Legacy Survey, a collaborative project of NRC and CNRS. Based on data products from observations made with ESO Telescopes at the La Silla Paranal Observatory under ESO programme IDs 294.A-5018, 294.A-5039 and 179.A-2005, and on data products produced by TERAPIX and the Cambridge Astronomy Survey Unit on behalf of the UltraVISTA consortium. The authors acknowledge the award of service time (SW2014b20) on the William Herschel Telescope (WHT). WHT and its service programme are operated on the island of La Palma by the Isaac Newton Group in the Spanish Observatorio del Roque de los Muchachos of the Instituto de Astrofisica de Canarias.

\bibliographystyle{apj}
\bibliography{references.bib}

\end{document}